%% file: StanfordBagleySolomon2015.tex
\documentclass[12pt]{article}
\usepackage{graphicx,color,geometry}
\usepackage{amsmath,amssymb,latexsym,dsfont,amsthm}
\usepackage[square,sort,comma,numbers]{natbib}
\pdfoutput=1

\usepackage{floatpag}
\usepackage{authblk}
\usepackage{placeins}
\usepackage{fancyvrb}
\usepackage[titletoc,title]{appendix}
\usepackage[utf8]{inputenc}
\usepackage[scaled=.8]{beramono}
\usepackage[T1]{fontenc}
\usepackage[hang,small,bf]{caption}

\floatpagestyle{empty}
\input{style.tex}

\newtheorem{defn}{Definition}
\newcommand{\mz}{\ensuremath{m/z}}
\newcommand{\il}{\ensuremath{{\triangledown}}}
\newcommand{\ir}{\ensuremath{{\vartriangle}}}
\definecolor{linkblue}{rgb}{0.192,0.494,0.675}
\newcommand{\webtext}[1]{\textcolor{linkblue}{\texttt{\footnotesize{#1}}}}

\newcommand\blfootnote[1]{%
  \begingroup
  \renewcommand\thefootnote{}\footnote{#1}%
  \addtocounter{footnote}{-1}%
  \endgroup
}

\title{Informed baseline subtraction of proteomic mass spectrometry data aided by a novel sliding window algorithm}
\date{}
\author[a,$\dagger$]{Tyman E. Stanford}
\author[ ]{Christopher J. Bagley}
\author[a]{Patty J. Solomon}
\affil[a]{\footnotesize{School of Mathematical Sciences, The University of Adelaide}}
\affil[$\dagger$]{To whom correspondence should be addressed: School of Mathematical Sciences, The University of Adelaide, Ingkarni Wardli Building, North Terrace, Adelaide 5005, Australia, Email: \texttt{tyman.stanford@adelaide.edu.au}}

\begin{document}
\bibliographystyle{unsrt}

\maketitle

\begin{abstract}
\footnotesize{
\hspace{-1cm} \textbf{\textsf{Background}} \hspace{0.4cm} 
Proteomic matrix-assisted laser desorption/ionisation (MALDI) linear time-of-flight (TOF) mass spectrometry (MS) may be used to produce protein profiles from biological samples with the aim of discovering biomarkers for disease or discrimination of disease states. 
The raw protein profiles suffer from several sources of bias or systematic variation, known as batch effects, which need to be removed before meaningful downstream analysis of the data can be undertaken. 
An early pre-processing step is baseline subtraction, which is the removal of non-peptide signal from the spectra.
Baseline subtraction is complicated by each spectrum having, on average, wider peaks for peptides with higher mass-to-charge ratios ({\mz}).
Additionally, the trial-and-error process of optimising the baseline subtraction input arguments is time-consuming and error-prone.
We present an analytical pipeline to overcome these current difficulties.
  
\hspace{-1cm} \textbf{\textsf{Methods}}  \hspace{0.4cm} 
Current best practice baseline subtraction is performed by partitioning the spectra into smaller regions.
The baseline subtraction method is then applied with constant and optimised input arguments within each region. 
We propose a new approach which transforms the {\mz}-axis to remove the relationship between peptide mass and peak width. 
Our preferred baseline subtraction method of the top-hat operator employs fast sliding window algorithms such as the line segment algorithm which cannot be applied to  unevenly spaced data. 
We have also developed a novel `continuous' line segment algorithm to efficiently operate on unevenly spaced data. 
To reduce the need for user input and the possibility of user error, we additionally present an input-free algorithm to estimate peak widths on the transformed {\mz} scale and thus the required sliding window widths for the top-hat operator. 
The methods are validated using six publicly available proteomic MS datasets.

\hspace{-1cm} \textbf{\textsf{Results}} \hspace{0.4cm} 
The automated baseline subtraction method was deployed on each dataset using six different {\mz}-axis transformations. 
The resulting baseline subtracted signal was compared to the gold-standard piecewise baseline subtracted signal. 
Optimality of the {\mz}-axis transformation when using the automated baseline subtraction pipeline was assessed quantitatively using the mean absolute scaled error (MASE). 
Several of the transformations investigated were able to reduce, if not entirely remove, the peak width and peak location relationship. 
The best performing transformations achieved automated baseline subtractions very similar to the gold-standard.
The proposed novel `continuous' line segment algorithm is shown to far outperform naive sliding window algorithms with regard to the computational time required, on both real and simulated unevenly spaced MALDI TOF-MS data. 
The improvement observed in the time required to compute baseline subtraction on the six MALDI TOF-MS datasets was at least four-fold and at least an order of magnitude on many simulated datasets.

\hspace{-1cm} \textbf{\textsf{Conclusions}}  \hspace{0.4cm} 
The new  pipeline presented here for performing baseline subtraction has a number of advantages over currently available methods. 
These advantages are: informed and data specific input arguments for baseline subtraction methods, the avoidance of time-intensive and subjective piecewise baseline subtraction, and the ability to automate baseline subtraction completely. 
Moreover, individual steps can be adopted as stand-alone routines. 
For example, the algorithm to automatically estimate peak widths can be used to dynamically calculate initial baseline subtraction method input arguments for subsequent user refinement for any given dataset. 
The proposed automated pipeline produced near-optimal baseline subtraction when compared to the current gold-standard of piecewise baseline subtraction method.
}

\end{abstract}


\section*{Background}

\subsection*{Discovery of protein biomarkers by mass spectrometry}

Protein biomarkers are proteins or protein fragments that serve as markers of a disease or condition biomarkers \cite{Albrethsen2007,Kulasingam2008} by virtue of their altered relative abundance in the disease state versus the healthy condition. 
Matrix-assisted laser desorption/ionisation (MALDI) linear time-of-flight (TOF) mass spectrometry (MS) is a widely used technology for biomarker discovery as it can create a representative profile of polypeptide expression from biological samples. 
These profiles are displayed as points of polypeptide abundance (intensity; the $y$-axis) for a range of mass-to-charge values ({\mz}; the $x$-axis). 
Each spectrum is an array of positive intensity values for discretely measured {\mz} values, but the profile is typically displayed on a continuous scale. 
MALDI TOF-MS spectra are typically limited to polypeptides less than 30 kilo Daltons although there is no theoretical upper limit \cite{Hortin2006}. 
Numerous biomarkers using MALDI TOF-MS have been identified to date \cite{Hortin2006,Croxatto2012}.\blfootnote{Abbrevations used: AMASE: average mean absolute scaled error;
CLSA: continuous line segment algorithm;
Da: Dalton;
EPCP: estimated peak coverage proportion; 
LOESS: Locally weighted scatterplot smoothing;
LSA: line segment algorithm;
MALDI: matrix-assisted laser desorption/ionization;
MASE: mean absolute scaled error;
MS: mass spectrometry;
MSE: mean squared error;
SNIP: sensitive nonlinear iterative peak;
TOF: time of flight.}

\begin{figure}[h!] \centering
  \includegraphics[width=0.8\textwidth]{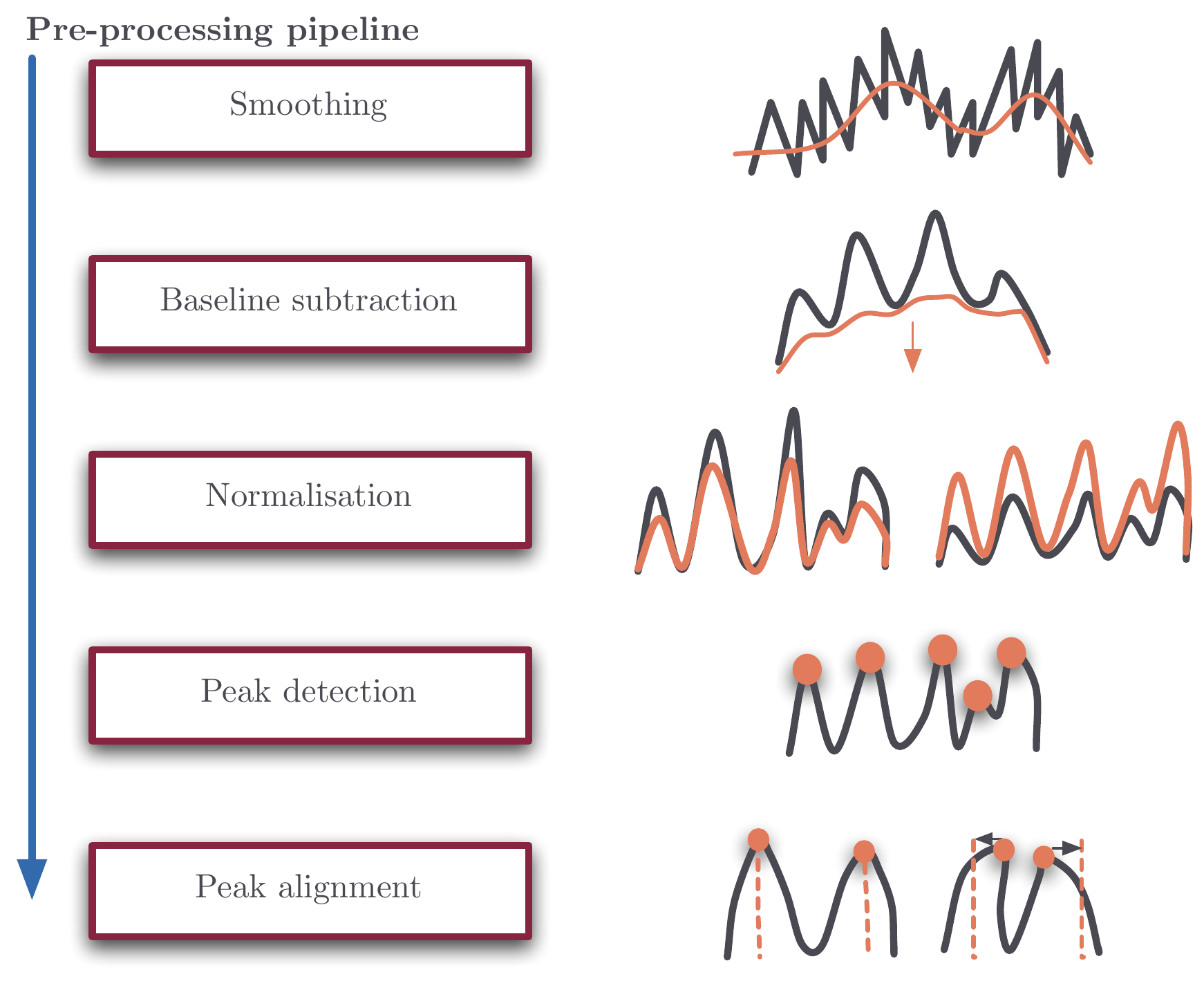}
  \caption{\textbf{The spectra pre-processing pipeline.}
      The steps, in order, required to successfully pre-process raw proteomic MALDI TOF-MS data.} \label{fig:pp}
      \end{figure}

Statistical analysis of the proteomic profiles for biomarker discovery cannot be undertaken without prior removal of noise and systematic bias present in the raw spectra. 
This removal is conducted through a series of steps known as  pre-processing. 
Pre-processing generally consists of five steps to remove false signal as set out in Figure~\ref{fig:pp}: signal smoothing, baseline subtraction, normalisation, peak detection and peak alignment. 
Signal smoothing and baseline subtraction are adjustments made to each spectrum individually (i.e., intra-spectrum pre-processing), 
while
normalisation and peak alignment (after peak detection) are adjustments made to make each spectrum within an experiment comparable (i.e., inter-spectrum pre-processing).

Signal smoothing is the first step in pre-processing the data and aims to remove instrument-derived noise in the data and stochastic variation in the spectrum signal. 
Baseline subtraction then follows, which is the removal of the estimated `bed' on which the spectral profile sits, composed of non-biological signal, e.g. chemical noise from ionised matrix.
Normalisation is the third step in pre-processing. This has the aim of making the observed signals proportionate over the experiment; to correct for instrument variability and sample-ionisation efficiency that will influence the number of peptide ions reaching the detector.
Peak detection is the fourth step, which is the detection of peak signal as peptide mass and intensity pairs. 
Finally, in the fifth step, the peaks are subject to peak alignment which adjusts for small drifts in {\mz} location which result from the calibration required for the TOF-MS system. This ensures that peptides common across spectra are recognised and compared at the same {\mz} value. 
Once the data have been pre-processed, analysis to detect potential biomarkers can be performed.

There are numerous freely available MS pre-processing packages. For example, in the \texttt{R} statistical software environment, \texttt{MALDIquant}, \texttt{PROcess} and \texttt{XCMS} are available \cite{R2014,Bioconductor2004,MALDIquant,Li2005,Smith2006}. 
Although we have set out the usual sequence of five data pre-processing steps, an optimal approach to pre-processing is not yet established and there is scope to improve current pre-processing methods and the order in which they are applied, to allow more reliable biomarker identification \cite{Stanford2015}. 
The present paper focuses on optimising methods for the baseline subtraction step of pre-processing of the raw spectra.

\subsubsection*{Baseline subtraction}

\begin{figure}[hp] \centering
  \includegraphics[width=\textwidth]{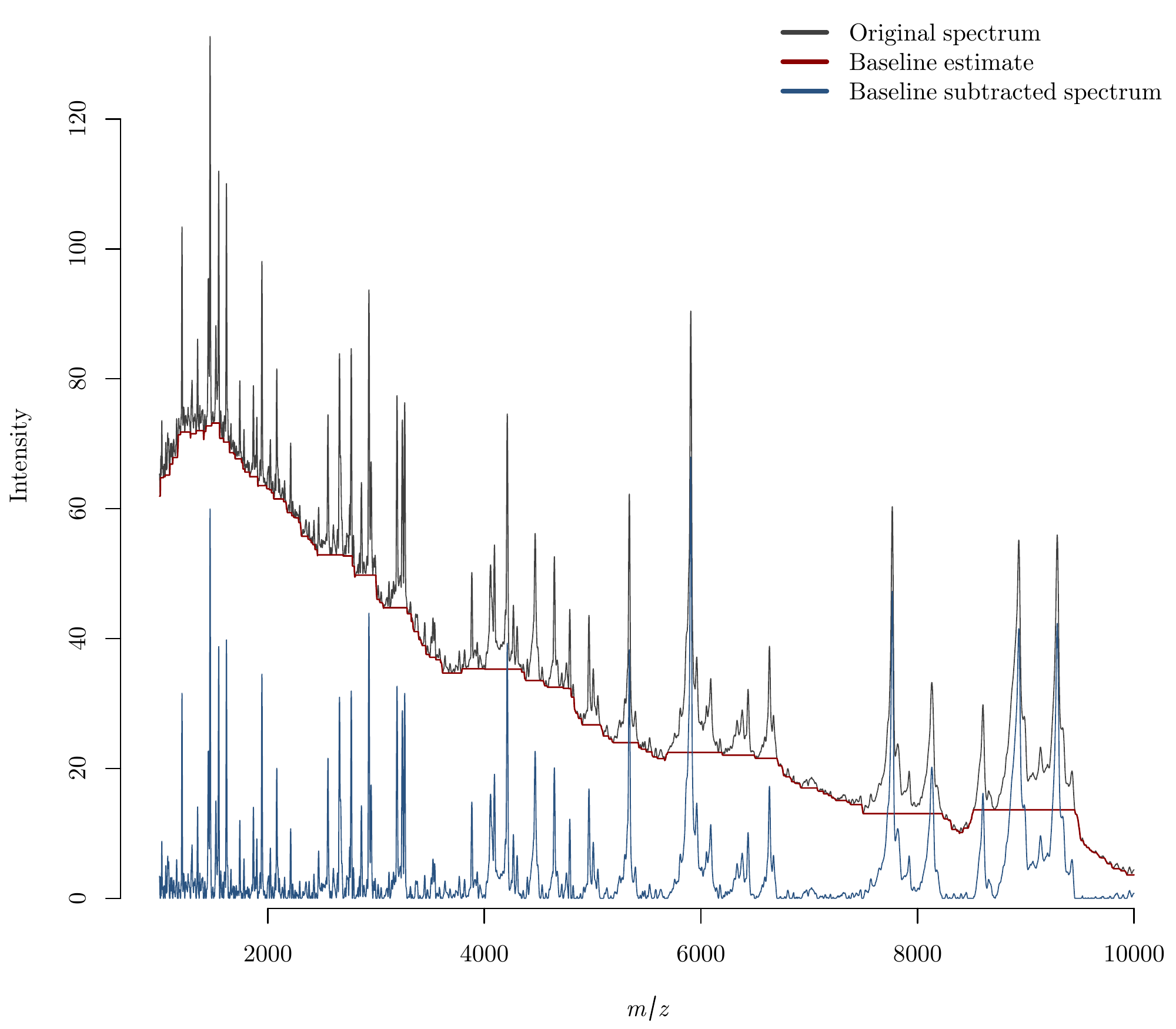}
  \caption{\textbf{Baseline subtraction of a proteomic MALDI-TOF mass spectrum.}
      A spectrum  from the Fiedler dataset: see `Data used' in the Methods section. The square root of the spectrum intensities was taken as a variance stabilisation measure and smoothing using Savitzky-Golay (half window size of 50) was applied prior to baseline subtraction.} \label{fig:blex}
      \end{figure}

The non-biological signal to be removed by baseline subtraction is often described as `chemical noise' which predominantly occurs at low mass values and may result from ionised matrix molecules \cite{Glish2003}.
An example of a MALDI TOF-MS spectrum, a baseline estimate and the resulting baseline subtracted spectrum are shown in Figure~\ref{fig:blex}. 
The spectrum in Figure~\ref{fig:blex} is a from the Fiedler dataset which is outlined in the `Data used' section in Methods. 
The pre-processing applied prior to the baseline subtraction involved taking the square root of the spectrum intensities (for variance stabilisation) and performing the first pre-processing step in smoothing using the Savitzky-Golay method with a half window size of 50 \cite{Savitzky1964}.

The baseline subtraction method discussed in the present paper utilises the \textit{top-hat} operator, which is an operator  defined in mathematical morphology. Mathematical morphology was originally proposed for two-dimensional image analysis then further developed for  image processing of microarray data images \cite{Yang2002,Mayer2005}. It has since been applied to MS data \cite{MALDIquant,Sauve2004,Kohlbacher2007,Lange2007,Sturm2008,Bauer2011}, and we describe the theory that is largely ignored when applied naively. The mathematical morphology definitions of an erosion, dilation, opening and top-hat provided below allow us to extend the current use of mathematical morphology in MS baseline subtraction.

The top-hat operator has some properties, i.e. it is a non-parametric and non-linear filter, which make it desirable for baseline subtraction. 
In particular, this suits the non-biological signal in MS spectra which may not follow a known functional form. 
Furthermore, the top-hat operator is computationally inexpensive compared with standard functional filters that require estimates of model parameters. 

Other algorithmic methods of baseline subtraction such as the sensitive nonlinear iterative peak (SNIP) algorithm \cite{Morhac2009,Ryan1988} provide an alternative to the top-hat operator. 
However, it will be shown in the Methods section that the top-hat operator can importantly be extended, using the mathematical theory underpinning it, for unevenly spaced data.

Standard methods of baseline subtraction estimate local minima (troughs) and fit either local regression (LOESS, Savitzky-Golay) or interpolate (splines) through these points \cite{Yang2009}. 
These methods require careful selection of the window size for detecting troughs, the polynomial order and the span of points for fitting the model, where applicable. 
Despite using optimised input arguments for these methods, they cannot guarantee a non-negative resultant signal. 
In fact, padded or removed signal in places of high curvature in the spectra may be produced.  
This can easily be envisioned by considering two local minimums and an adjacent point to one of the local minimums that lies between both.
There is no property that stops the adjacent point lying below an interpolation of the two minimums, especially where there exists a large difference between the values of the local minimums.

\subsection*{Morphological image analysis and theory}

The core concepts in mathematical morphology required to apply the top-hat operator are presented below. 
The definitions of a morphological \textit{structuring element}, \textit{dilation}, \textit{erosion}, \textit{opening} and \textit{top-hat} can also be found in \cite{Dougherty1992,Soille1999,Gil2002,VanDroogenbroeck2005}.

A structuring element (SE) is a small set that acts on given data or images.
For linear TOF-MS data, a SE is simply a one-dimensional line-segment, or window, passed over the vector of spectral intensities. 
In the context of morphological image analysis, the SEs used are centred (the median SE value is 0), symmetric (the SE behaves the same either side of the centre) and flat (SE is or the same dimension as the data). 
Non-flat SEs are not ideal for the current application as they require a known function or weightings to be applied within the sliding window. 
\begin{defn} \label{def1}
For the sets $X \subset \mathds{Z}^p$ and $B \subset \mathds{Z}^{p}$, $p \in  \mathds{Z}^{+}$, and the function $f$  defined over $X$, the erosion of $X$ by $B$ is defined as, 
\begin{eqnarray*}
\epsilon_{B} \left( f \right) \left( x \right)  & := & \left( f \ominus B \right) \left( x \right) \\
& := & \inf_{b \in B} f \left( x + b \right),
\end{eqnarray*} 
for each element $x$ in $X$. The dilation is similarly defined, 
\begin{eqnarray*} 
\delta_{B} \left( f \right) \left( x \right)   &:= & \left( f \oplus B \right) \left( x \right) \\ 
&:=& \sup_{b \in B} f \left( x + b \right).
 \end{eqnarray*} 
\end{defn}

Erosions and dilations can be thought of as rolling minimums and maximums, respectively, over the spectral values.
Sometimes the sets $X$ and $B$ in Defintion~\ref{def1} are defined over $\mathds{R}^p$ \cite{Dougherty1992,VanDroogenbroeck2005} but this is rarely implemented for data other than $X\subset\mathds{Z}^p$ in practice.

\begin{defn}
The application of a morphological erosion followed by a morphological dilation to a set $X$ is the morphological opening,
\begin{eqnarray*}
\omega_{B} \left( f \right) \left( x \right)  & := &\delta_{B}  \left( \epsilon_{B}  \left( f \right) \right)  \left( x \right)   \\
   & := &  \left(\left( f \ominus B \right) \oplus B \right)\left( x \right) .
\end{eqnarray*}
\end{defn}

In the context of linear TOF-MS data, a morphological opening is a non-linear estimation of background signal of the one-dimensional spectrum on $X$. 
The opening has the desirable property that it is never returns values greater  than the observed signal, i.e. $\omega_{B} \leq f \; \forall \; x \in X$. 

\begin{defn} The top-hat operator is defined as the removal of the opening from the original signal $f$,
\begin{eqnarray*}
\tau_{B} \left( f \right) \left( x \right)  & := &  f \left( x \right) - \omega_{B}  \left( f \right)\left( x \right)    .
\end{eqnarray*}
\end{defn}

The result of applying the top-hat operator to proteomic TOF-MS is the estimation of the true signal by removing the estimated background signal from $f$ on $X$. 
Because of the $\omega_{B} \left( f \right) \leq f$  $\left( \forall \; x \right)$ property of morphological openings, the top-hat operator provides a background estimate and removal without risk of creating negative signal, since it is a physical impossibility of the system. 
Such properties cannot be guaranteed by local regression of local minima.

\subsubsection*{Example of the top-hat operator}

To illustrate the morphological operators that have been defined, we consider a simple example. Let $f = \left\{ a_x \right\}_{x=1}^{13}$ be a series and define a flat SE, $B= \left\{ b_j \right\}_{j=1}^{5} = \left\{ -2,-1,0,1,2 \right\}$ with 
\[
f \left( x \right) =
\begin{cases} 
a_1 & \text{if $x<1$}\\
a_x &\text{if $x=1,2, \ldots, 13$} \\
a_{13} & \text{if $x>13$,}
\end{cases}
\]
where
\[
\left\{ a_x \right\}=\left\{ \begin{array}{ccccccccccccc}
6 & 11 & 12 & 14 & 7 & 10 & 13 & 9 & 12 & 15 & 8 & 11 & 10 \end{array} \right\}.
\]

The erosion at $x=4$ is calculated,
\begin{eqnarray*}
\epsilon_{B} \left( f \right) \left( 4 \right)  
&=& \inf_{b \in B} f \left( 4 + b \right) \\
&=& \inf \left\{ 11 , 12 , 14 , 7 , 10\right\} 
= 7.
\end{eqnarray*}

Given the erosions for $x=2,3,4,5,6$ are $6,6,7,7,7$, respectively, the morphological opening at $x=4$ is
\begin{eqnarray*}
\omega_{B} \left( f \right) \left( 4 \right)  
&=& \delta_{B}  \left( \epsilon_{B}  \left( f \right) \right) \left( 4 \right) \\
&=& \sup_{b' \in B}
\left\{
  \epsilon_{B}  \left( f \right) \left( 4 +b'  \right)
 \right\} \\
&=& \sup
\left\{
 \inf_{b \in B} f \left( 2 + b \right),
 \inf_{b \in B} f \left( 3 + b \right),
 \inf_{b \in B} f \left( 4 + b \right), \right. \\
&& \hphantom{\inf_{b \in B} f \left( 2 + b \right),
 \inf_{b \in B} f \left( 3 + b \right),} 
 \left. \inf_{b \in B} f \left( 5 + b \right),
 \inf_{b \in B} f \left( 6 + b \right)
 \right\} \\
&=& \sup \left\{ 6,6,7,7,7 \right\} = 7.
\end{eqnarray*}

Therefore, the top-hat operator result for $x=4$ is
\begin{eqnarray*}
\tau_{B} \left( f \right) \left( 4 \right)   =   f \left( 4 \right) - \omega_{B}  \left( f \right)\left( 4 \right)=14-7=7    .
\end{eqnarray*}

  \begin{figure}[h!] \centering
  \includegraphics[width=0.8\textwidth]{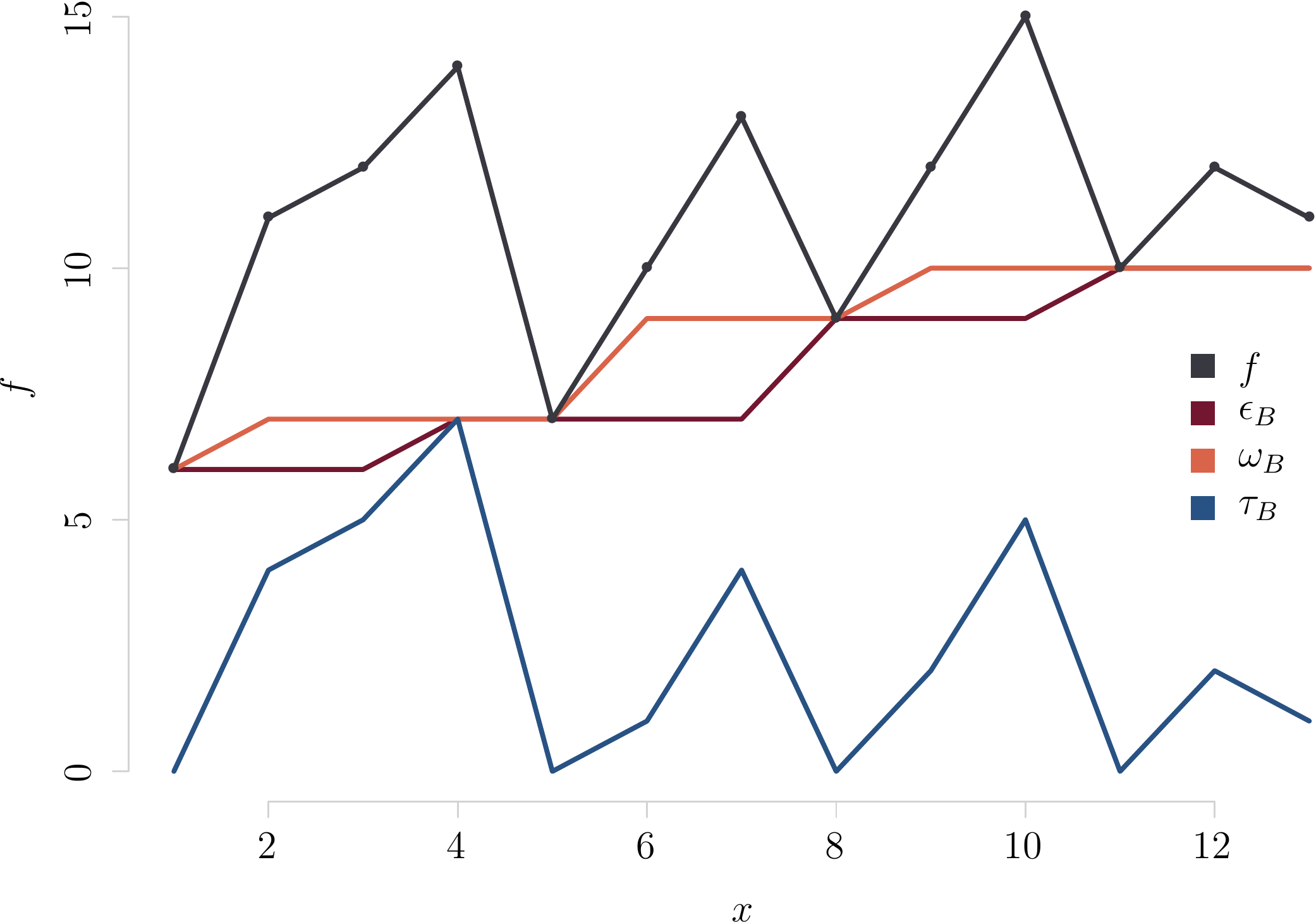}
  \caption{\textbf{Baseline subtraction on an example spectrum using the top-hat operator (see main text for details):}
     an demonstration of the erosion, opening and top-hat operators ($\epsilon_{B}$, $\omega_{B}$ and $\tau_{B}$, respectively) on a set $f$.} \label{fig:thex}
      \end{figure}

The operations with $\epsilon_{B}$, $\omega_{B}$ and $\tau_{B}$ using the flat SE, $B= \left\{ -2,-1,0,1,2 \right\}$, on the entire signal $f(x)$ can be observed in Figure~\ref{fig:thex}.

\subsubsection*{Current application of the top-hat operator to linear TOF-MS}

A naive algorithmic application of an erosion to spectral intensities simply requires a traversal of each point, where the minimum value within a window over that point is the resulting erosion. 
The process is performed similarly  for a dilation. 
However, erosions and dilations can be calculated more efficiently with the line segment algorithm (LSA) \cite{vanHerk1992,Gil1993}. 
Application of the LSA is mainly seen in medical imaging and analysis \cite{vanHerk1998,Heneghan2002}. The \texttt{R} package \texttt{MALDIquant} and \texttt{OpenMS} use this algorithm in their implementation of the top-hat operator.

When applying the top-hat operator to a spectrum, the SE needs to be chosen carefully. 
In particular, the following need to be considered.
\begin{enumerate}
\item If a SE is too large, then it will be too conservative and leave false signal.
\item If a SE is too small, it will result in under-cut peaks and remove valid signal.
\item The mean peak width increases further along the {\mz}-axis \cite{Zhang2010}. The baseline subtraction needs to be performed in a piecewise manner, otherwise the above issues 1 and 2 will occur.
\end{enumerate}

Despite the simplicity of the top-hat operator compared to functional alternatives, piecewise baseline subtraction is still required. 
In fact, piecewise baseline subtraction should be applied for any method that implicitly assumes peak width remains constant, such as local regression, interpolating splines or the SNIP algorithm.

The SE size used for the top-hat operator needs to be of equivalent window size to each spectrum's peak widths, or greater, to ensure the top-hat operator does not `undercut' peak intensities. 
The piecewise baseline subtraction involves determining subsections of the {\mz}-axis, where fixed SE widths (in the number of {\mz} points) in each section are appropriate, or the equivalent input arguments for other baseline methods. 
Smaller SEs will be chosen corresponding to lower {\mz} values and larger SEs will be used corresponding to larger {\mz} values.

\begin{figure}[hp] \centering
  \includegraphics[width=\textwidth]{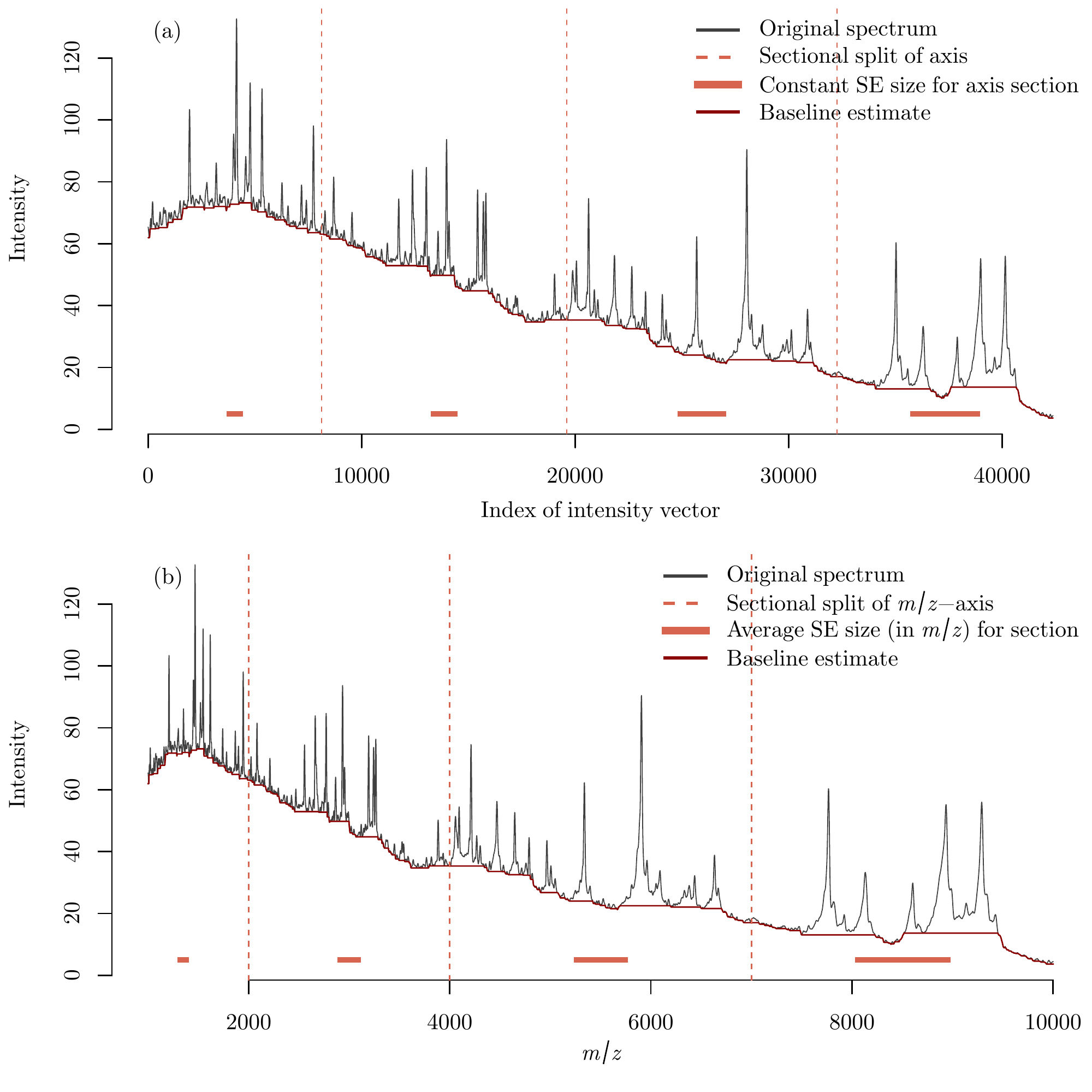}
  \caption{\textbf{Piecewise baseline subtraction of a proteomic MALDI-TOF mass spectrum from the Fiedler dataset using the top-hat operator.}
 Because the SE is provided as a number of {\mz} points and {\mz} values increase in distance along the axis, the SE size is not constant within sub-intervals of the {\mz}-axis. } \label{fig:blpw}
      \end{figure}

Figure~\ref{fig:blpw}(a) illustrates a spectrum from the Fiedler data separated into four roughly equal segments based on the number of intensity values.
When applying the top-hat operator, the SE size is a constant number of intensity values within each piecewise section of the axis.
The SE sizes selected in Figure~\ref{fig:blpw}(a) were made by visual inspection and trial-and-error. 
Figure~\ref{fig:blpw}(b) depicts the same spectrum as Figure~\ref{fig:blpw}(a) but the $x$-axis is in terms of {\mz} location.
On this {\mz}-axis, the SE size increases along the {\mz}-axis within each piecewise segment simply by virte of the distances between {\mz} points increasing, even though the same window size is being used in terms of the number of intensity values. 
However, the increasing coverage in {\mz} units across the {\mz}-axis is not proportional to the increase in peak widths.
Figures~\ref{fig:blpw}(a) and~(b) demonstrate that there is not a constant number of intensity values for the SE across the entire {\mz}-axis that could avoid conservative baseline estimates (1) or under-cut peaks (2) or even both.

\subsubsection*{Improving baseline subtraction}

Prior to pre-processing MALDI TOF-MS data, a log or square root transformation of the intensity axis is usually performed as a variance stabilisation measure but no such transformation is made to the {\mz}-axis.
If an appropriate {\mz} transformation could be made however, piecewise pre-processing of the spectra for the baseline subtraction step (and potentially for other pre-processing steps) could be avoided.
Additionally, the default arguments such as window size in software to perform baseline subtraction are statically defined.
Uninformed default arguments such as these are highly likely to need modification for successful baseline subtraction, as spectra attributes vary from one experiment to another.  Dynamic default arguments that are informed by the data would be an advantage in saving both user time and minimising user error.

\section*{Methods}

\subsection*{A pipeline to achieve automated baseline subtraction}

\begin{figure}[hp] \centering
  \includegraphics[width=0.8\textwidth]{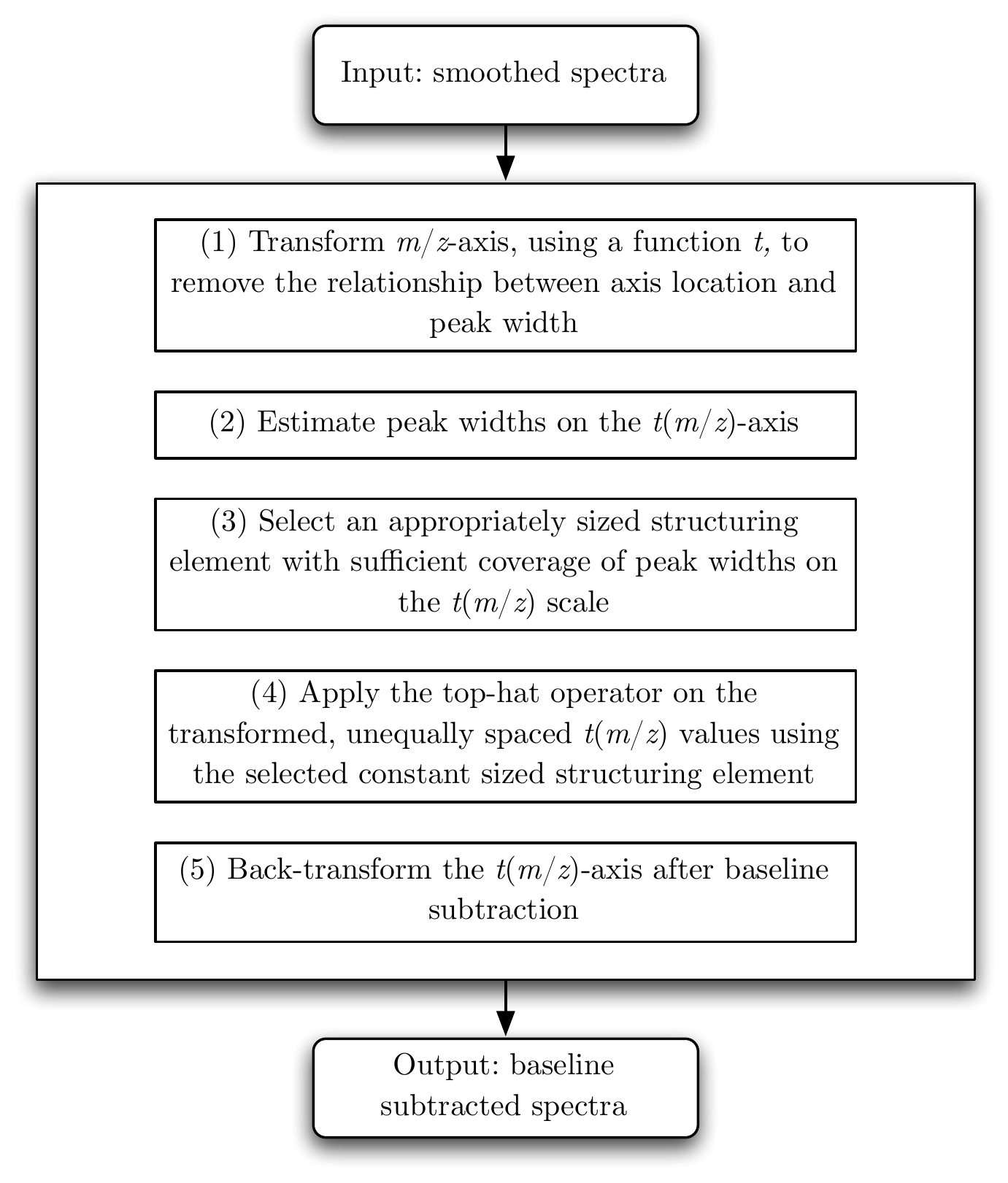}
  \caption{\textbf{The proposed baseline subtraction pipeline}:
      five steps for automated baseline subtraction.} \label{fig:pipe}
      \end{figure}

The pipeline shown in Figure~\ref{fig:pipe} can be employed to automate the baseline subtraction step.
The first step of the pipeline requires a suitable transformation of the {\mz}-axis. 
If such a transformation of the {\mz}-axis can be made, a piecewise approach is not required as a constant-sized SE can be used over the entire spectrum.  
A log-type transform that expands the low {\mz} values and contracts the high {\mz} values is required. Once a suitable transformation is  found, a top-hat operator defined over non-evenly spaced real values (i.e. $X\subset\mathds{R}^p$, as opposed to integer values) can be used at the baseline subtraction step. The implementation requires a minimum and a maximum sliding window algorithm for unevenly spaced data which means the LSA cannot be used. 
Naive algorithms are available; however, here we present a novel sliding window algorithm that we show outperforms naive sliding window algorithms by avoiding repeated minimum (or maximum) calculations for common points in successive sliding windows.
However, a SE size does need to be selected.
This can be implemented by firstly estimating peak widths, then selecting a SE size that covers a sufficient proportion of the estimated peak widths.
The process of estimating peak widths can be automated without user input and our recommend approach is presented here.
The final step in the baseline subtraction pipeline is simply the (reverse) transformation back to the original {\mz} scale.

The new pipeline to perform baseline subtraction of MALDI TOF-MS data presented in Figure~\ref{fig:pipe}  has two major advantages when compared to standard methods.
\begin{itemize}
\item Firstly, the  pipeline  automates the baseline subtraction step, that is otherwise conducted in a  piecewise manner. This eliminates  the need for user input and time-consuming calibration by observation. Automation of the baseline subtraction step also minimises the potential for user error and the time required to assess the input arguments for optimality.
\item Secondly, the novel algorithm that is computationally less expensive than a naive minimum or maximum sliding window algorithm to perform the top-hat operation on unevenly spaced data, presented here,   further minimises the computational time burden of baseline subtraction. 
\end{itemize}

Fields of application outside of bioinformatics that encounter unevenly spaced data are  also likely to find this algorithm useful in practice. Other names for unevenly spaced data include unevenly sampled, non-equispaced, non-uniform, inhomogeneous, irregularly sampled or non-synchronous data. 
Such data occur in various fields including, but not limited to; financial time-series, geologic time-series, astrophysics and medical imaging \cite{Greengard2004,Lo1990,Aris2005,Schulz2002,Deeming1975,Scargle1982,Bourgeois2001}.
 Analysis and processing of unevenly spaced data is an ongoing field of research, as most methods for analysis assume equally spaced data.

\subsection*{Data used}

Six proteomic MS datasets from previously published studies were used to validate the methods presented. 

\begin{description}
\item[Fiedler data:] Urine samples were taken from 10 healthy women and 10 healthy men and peptides were separated  using magnetic beads (fractionation). The fractionated samples were then subject to MALDI TOF-MS \cite{Fiedler2007}. A subset of the MALDI TOF-MS data is freely available in the \texttt{R} package \texttt{MALDIquant} \cite{MALDIquant} and is the dataset used here. The spectra are observed over the range of values 1,000-10,000~{\mz}. 
\item[Yildiz data:] As described in \cite{Yildiz2007}, sera were collected from 142 lung cancer patients and 146 healthy controls to find relevant biomarkers. The serum samples were subject to MALDI TOF-MS without magnetic bead separation. The spectra are observed over the range of values 3,000-20,000~{\mz}. 
\item[Wu data:] MALDI TOF-MS data were generated from sera, as described in \cite{Wu2003,Yu2006}, with the aim of differentiating between 47 ovarian and 42 control subjects. The spectra are observed over the range of values 800-3,500~{\mz} using Reflectron mode which resolve peptide peaks into their isotopomers. 
\item[Adam data:] Surface-enhanced laser desorption/ionization (SELDI) TOF-MS data from 326 serum samples from subjects classified as prostate cancer, benign hyperplasia or control \cite{Adam2002}. While SELDI has been found to be less sensitive than MALDI, samples do not require fractionation before applying MS. The data analysed here are limited to the range 2,000-15,000~{\mz} as peptide signals beyond this range are sparse.
\item[Taguchi data:] The dataset available was first described in \cite{Taguchi2007} but is available as a supplement for \cite{Li2011}. The data are 210 serum-derived MALDI TOF mass spectra from 70 subjects with non-small-cell lung cancer with the aim of predicting response to treatment. The data observed cover the 2,000-70,000~{\mz} range.
\item[Mantini data:] The data in this study were produced using MALDI TOF-MS from purified samples containing equine myoglobin and cytochrome C \cite{Mantini2007}. A total of 30 spectra are available in the range 5,000-22,000~{\mz}.
\end{description}

\subsection*{Transformation of the {\mz}-axis}

The proposed pipeline for baseline subtraction requires a suitable transformation of the {\mz}-axis as the first step.
In this section we investigate potential transformations, that will be assessed  quantitatively for their suitability.

It has previously been suggested that peak width is roughly proportional to peak location on the TOF-axis \cite{Siuzdak2006,House2011} and that therefore peak width is proportional to the square of the {\mz} location. 
This was not in fact observed for any of the datasets analysed in the present study. 
Table~\ref{tab:trans} sets out the shortlist of suitable transformations, $t_0$-$t_5$, that are  considered appropriate for application here.

\begin{table}[h!]
\centering
\caption{The transforms, $t_i$, of the {\mz}-axis trialled to produce a roughly uniform distribution of peak widths across the $t_i\left({\mz}\right)$-axis.} \label{tab:trans}
\begin{tabular}{lr}
\hline
Label & Transform \\
  \hline
  $t_0\left(x\right)$ & $x$ \\
  $t_1\left(x\right)$ & $-1000x^{-1}$ \\
  $t_2\left(x\right)$ & $x^{1/4}$ \\
  $t_3\left(x\right)$ & $\ln x$ \\
  $t_4\left(x\right)$ & $-1000 \left(\ln x\right)^{-1}$ \\
  $t_5\left(x\right)$ & $-1000x^{-1/4}$  \\ 
  \hline
\end{tabular}
\end{table}

\begin{figure}[hp] \centering
  \includegraphics[width=\textwidth]{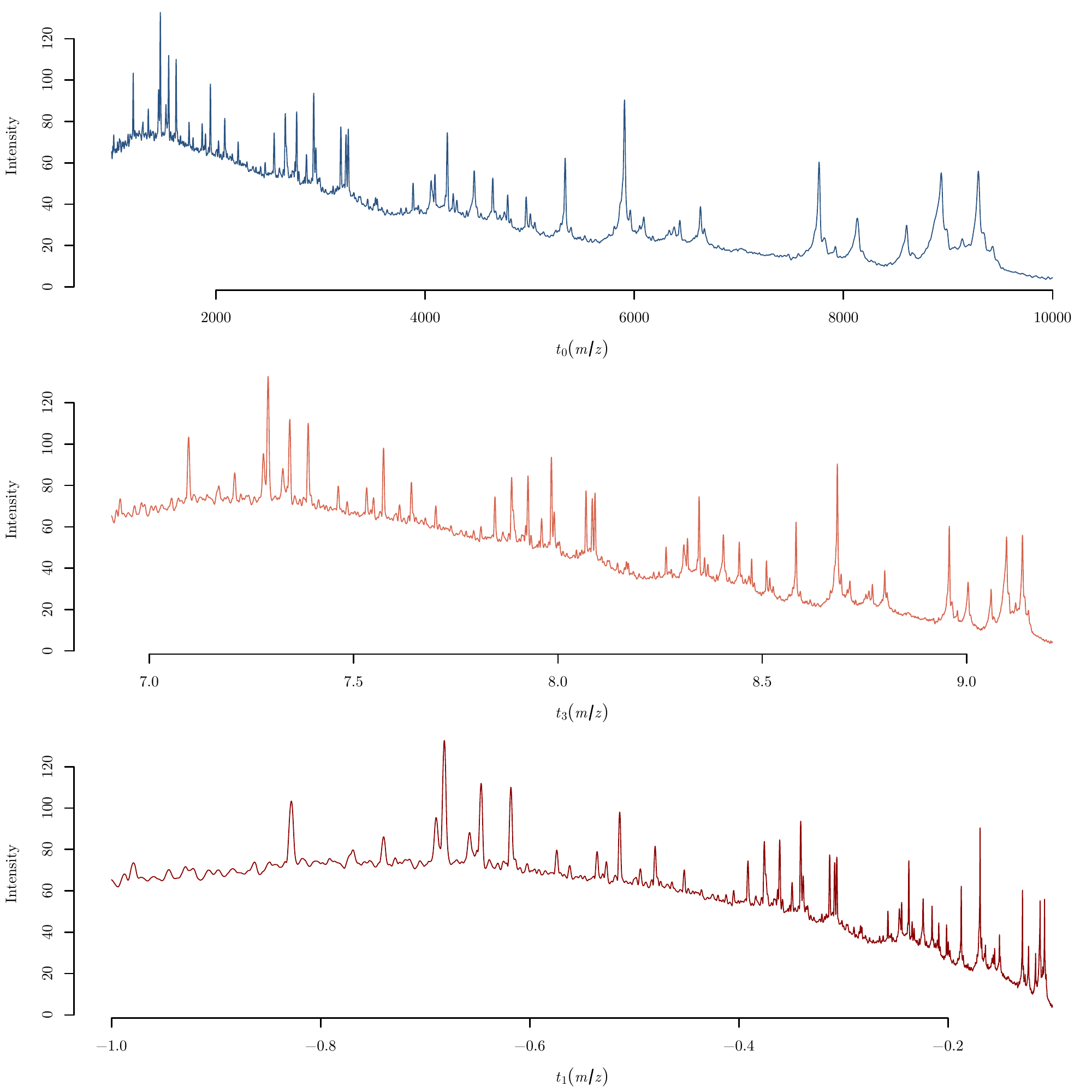}
  \caption{\textbf{Transformations of the {\mz}-axis:}
      Three different {\mz}-axis transformations (see Table 1) for the Fiedler spectrum shown in Figures~\ref{fig:blex} and~\ref{fig:blpw}.} \label{fivetrans}
      \end{figure}

To illustrate the role of the transformation, Figure~\ref{fivetrans} shows a spectrum from the Fiedler dataset on the original {\mz}-axis ($t_0$) for transformations $t_1$ and $t_3$. 
The effect of $t_3$, when compared to the original {\mz}-axis, is an expansion of smaller mass peak widths and the contraction of higher mass peak widths. 
However, visually it can be seen that higher mass peaks have larger peak widths on average even under the $t_3$ transformation. The $t_1$ transformation further shifts low {\mz} values across the transformed axis and contracts {\mz} values at the high end of {\mz}-axis. 
Potentially, the $t_1$ transformation creates larger peak widths for smaller {\mz} values than high {\mz} values so as to produce peak widths that decrease on average across the transformed axis. 
The effect of the six transformation functions, $t_0$-$t_5$, on a spectrum from each of the six datasets is available in Appendix~\ref{appa}.

\subsection*{Obtaining approximate peak widths prior to baseline subtraction}

Peak widths can be obtained at the peak detection step (step four of pre-processing) but such information is not generally known prior to the second pre-processing step of baseline subtraction.
To determine the constant SE size to be passed over the transformed {\mz}-axis, peak widths need to be estimated.
An algorithm to estimate peak widths from the data was created here for this purpose.

The algorithm below to estimate the peak widths within spectra takes the previously smoothed spectra on the transformed {\mz}-axis as the input and is performed as follows.
\begin{itemize}
\item For each spectrum, the lower convex hull of the two-dimensional set of spectrum points is used to determine an approximate baseline for each spectrum. 
\item The longest segment of the lower convex hull is then halved, with the two sets of points created by this split  subject to a new lower convex hull calculation. 
\item The newly calculated lower convex hull points for the two set of points are then added to the original set of lower convex hull points to improve the approximate baseline calculation. 
\item This is repeated $r-1$ more times to produce an approximate estimated baseline.
\item The approximate baseline is then removed and median intensity is then calculated for the resulting spectrum.
\item Intensities above the median value are treated as points along a peak.
\item The consecutive points above the median value are the estimated peak widths.
\end{itemize}

The above algorithm is crude and could not be used for reliable baseline subtraction. However, estimated peak widths are easily extracted using this method and can be used within the proposed automated baseline subtraction pipeline. 
 
\begin{figure}[hp] \centering
  \includegraphics[width=\textwidth]{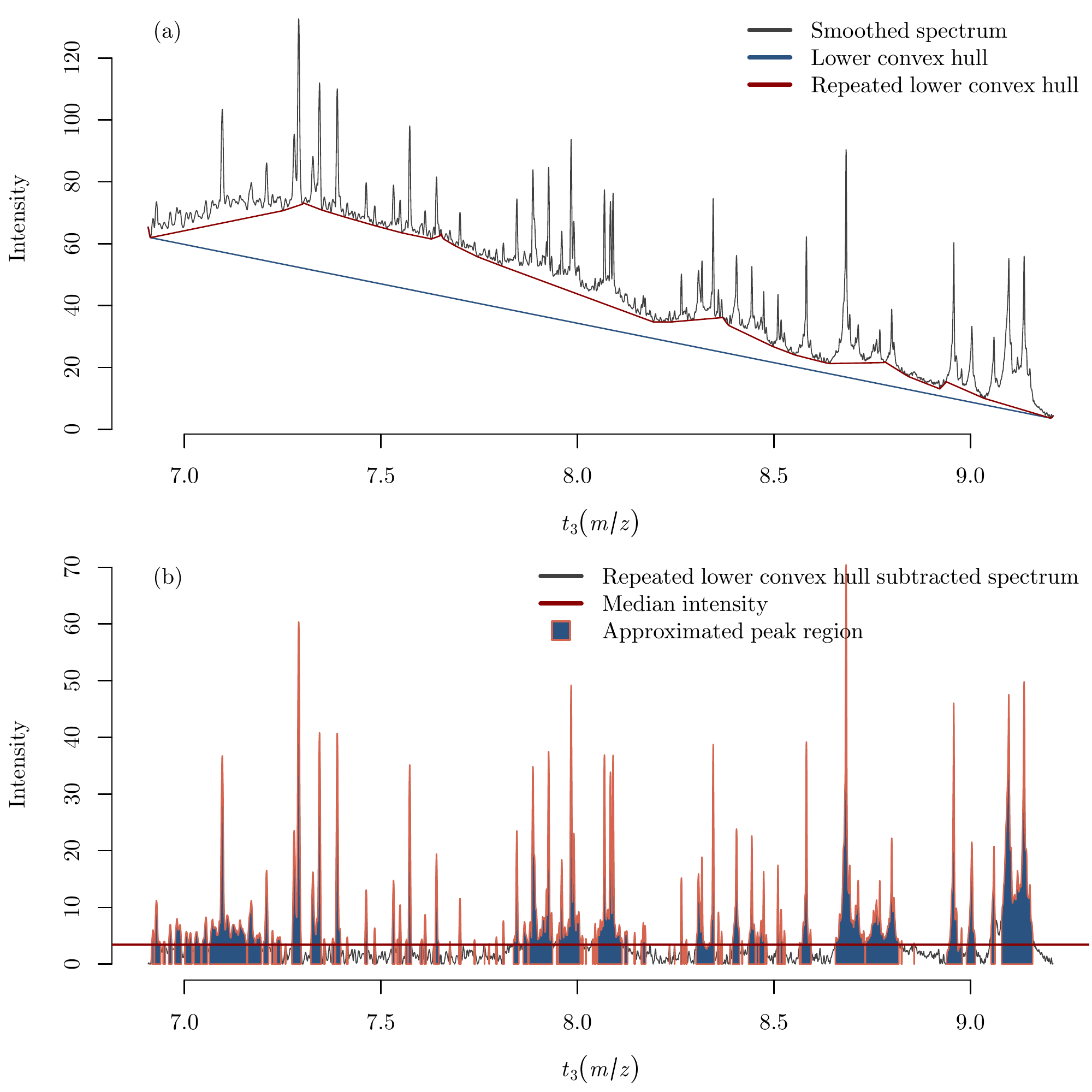}
  \caption{\textbf{Illustrated algorithm to determine approximate peak widths prior to baseline subtraction.}
      (a) A Fiedler spectrum with the lower convex hull and repeated ($r=5$) lower convex hulls shown. (b) The lower convex hull subtracted spectrum with median intensity of the spectrum depicted. Points above the median are considered peaks, which have been filled with colour.} \label{fig:chull}
      \end{figure}

A reasonable number of lower convex hull iterations of $r=5$ produced sensible results on the six datasets used. 
By specifying a value of $r$, this method to estimate peak widths is fully automated.
It provided enough alterations to the original lower convex hull to satisfactorily remove the residual baseline on concave smoothed spectra while not applying too many alterations so as to create midpoints along the longest segments which create lower convex hulls ending a peak vertices and therefore removing them. 
However, a missed peak or two per spectrum is not an issue as dozens of peaks are identified per spectrum. 
Figure~\ref{fig:chull} depicts this process, on a single spectrum.

The algorithm presented above attempts to automatically find peak widths without user input. 
We outline this automated procedure in the Methods section as it is not the focus of this paper, and may be substituted with any peak region finding method that requires no user input; such is the modularity of the pipeline shown in Figure~\ref{fig:pipe}.
There exist other methods to estimate peak widths (regions), such as that found in \cite{Morhac2009}, but they require previous knowledge of likely peak widths and are therefore not a baseline subtraction method that can be automated. 
\subsection*{Selecting a SE size and applying the top-hat operator in the transformed space}

Point three of Figure~\ref{fig:pipe} requires a choice of SE size. 
This can be chosen from the estimated peak widths found using the algorithm presented in previous section. 
The aim is to select a SE of sufficient size to not undercut peaks; such a SE size roughly translates to the maximum of the peak widths.
However, there is likely to be a SE size smaller than the maximum estimated peak width but much greater than the minimum estimated peak width that performs optimally.
Given a set of estimated peak widths for all spectra in an experiment and a SE size, we define the proportion of peak widths that are estimated to be the SE size or smaller as the estimated peak coverage proportion (EPCP) .

\begin{figure}[h!] \centering
\includegraphics[width=0.8\textwidth]{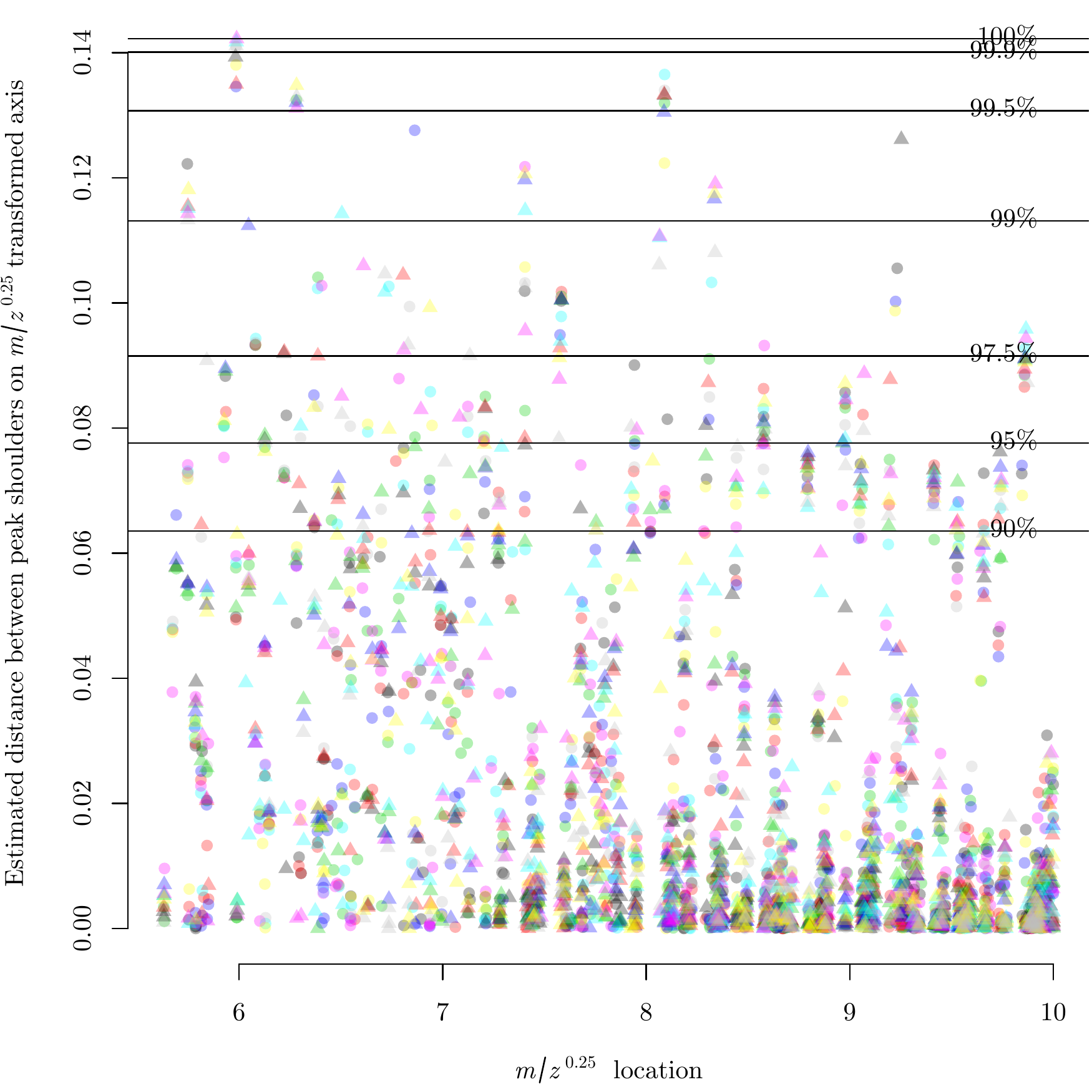}
  \caption{\textbf{Estimated peak widths on the $t_2\left(  m/z  \right)$ scale.}
      Peak widths in the Fiedler data with the {\mz}-axis transformed by the quartic root. The horizontal lines denote the proportion of peak widths that lie below it.} \label{pwtrans} 
\end{figure}

Figure~\ref{pwtrans} represents the estimated peak widths for the 16 spectra in the Fiedler dataset on the $t_2\left( {\mz} \right)$ scale, where peak regions are found using a repeated lower convex hull algorithm presented previously.

We trial different SE sizes corresponding to different EPCP values in the hope an optimal EPCP value for each of the six datasets we utilise can be found. 
A SE size that fully covers 97.5\% of detected estimated peak widths (EPCP of 0.975) for example, could yield optimised baseline subtraction. 

Both the EPCP and {\mz}-axis transformation are variables that are explored in the Results section, to find an empirically optimal combination.
Optimality of the automated baseline subtraction can be assessed by calculating the minimum value of an error metric relative to a gold-standard baseline subtraction, given a set of EPCP values and transformation functions. 
The metric used to compare the automated baseline subtraction to the gold-standard is outlined in the next section and the modified algorithm to perform top-hat baseline subtraction on the unevenly spaced and transformed {\mz}-axis is provided in the section after that.

\subsection*{Comparison of proposed methods to the gold-standard}

Piecewise, top-hat baseline subtracted spectra were used as the gold-standard baseline subtracted spectra. 
The SE sizes for each piecewise segment along the {\mz}-axis were selected using trial-and-error to produce the best baseline subtraction as determined visual inspection. 
These baseline subtracted, gold-standard spectra were produced prior to the automated baseline subtraction methods being applied.

Mean absolute scaled error (MASE \cite{Hyndman2006}) was selected to be the error metric of the automatically baselined spectra for a given transformation and EPCP, when compared to the gold-standard baseline subtracted spectra. 
Because the MALDI TOF mass spectra intensities are on arbitrary scales prior to normalisation, it is important to use a metric that is scale free, in order to be able to compare results between spectra from different experiments. 
MASE also avoids many degeneracy issues of other relative error metrics with zero denominators.
Baseline subtracted spectra will have many zero values where no signal is present. 
Other metrics such as mean squared error (MSE) were considered (which did not change the selection of the optimal transform and EPCP) however the ability to compare the error with other data is not possible and some sort of normalisation or weighting of spectra is required to ensure the MSE, say, of selected spectra do not dominate the result.

Let $\tau^{*}_j$ denote the intensity at $x_j$ of a gold-standard baseline subtracted spectrum $\tau_{B^{*}} \left( x_j \right)$  and  $\tau_j$ denote an automated baseline subtracted spectrum $\tau_B \left( x_j \right)$. 
The MASE is calculated as 
\[
\text{MASE} = 
\text{mean} 
\left(
\left\{ 
 \frac{
  | \tau^{*}_j- \tau_j | 
  }{ 
  \frac{1}{n-1} \sum_{i=2}^n  | \tau^{*}_i- \tau^{*}_{i-1} | 
  } 
\right\}_{j=1,2,\hdots,n} 
\right).
\]

For each of the six datasets, there are $N$ spectra to be compared.  Let AMASE be the average MASE value of the $N$ baseline subtracted spectra, then
\[
\text{AMASE}=\frac{1}{N}\sum_{\ell=1}^N \text{MASE}_{\ell}.
\]

\subsection*{The `continuous' line segment algorithm}

A novel algorithm is proposed here that can be applied to the unevenly spaced values of the transformed {\mz}-axis using a constant SE width. 
This algorithm, which we name the `continuous' line segment algorithm (CLSA), requires fewer computations per element than current rolling maximum and minimum algorithms on unevenly spaced data \cite{Eckner2012}.

\begin{figure}[hp] \centering
\includegraphics[width=\textwidth]{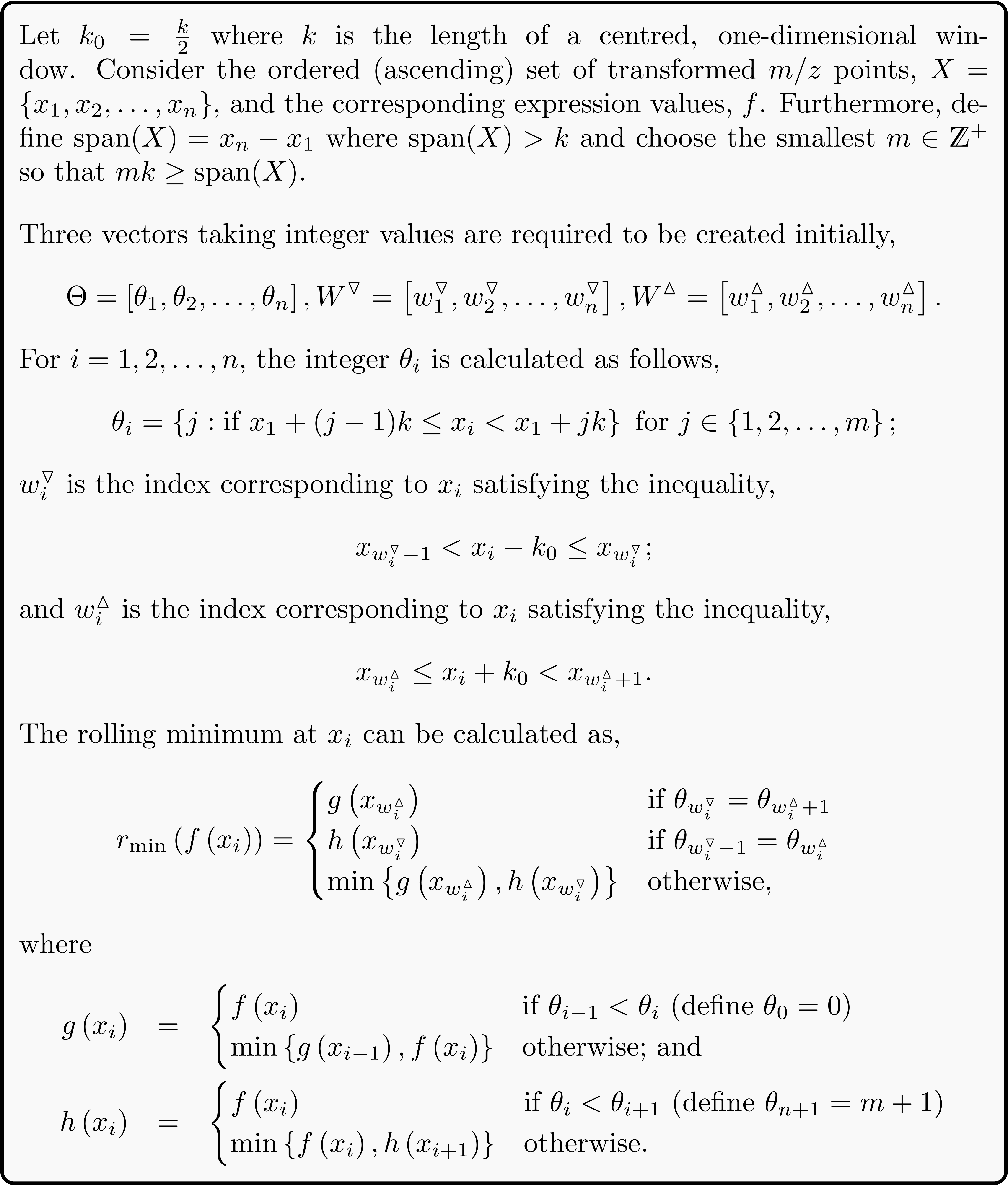}
\caption{\textbf{The continuous line segment algorithm (CLSA).}} \label{fig:clsaalg}
\end{figure}

Consider the case where values in $X$ are not evenly spaced,  and $X \subset \mathds{R}$, as opposed to $X=\left\{ 1,2,\ldots,n \right\}$, such as proteomic spectra on a transformed $t\left({\mz}\right)$-axis. 
Figure~\ref{fig:clsaalg} outlines the CLSA as a rolling minimum algorithm that can be trivially converted to a rolling maximum algorithm by finding the rolling minimum of $-f$ and returning the negative values of the result.

In effect, the CLSA creates $m$ blocks using the $\theta_i$ relating to the corresponding $x_i$:
\begin{eqnarray*}
 \theta_1, \theta_2,\ldots, \theta_{b_1}  &=    1   &  \text{where } x_1, x_2,\ldots, x_{b_1} \in \left[x_1,x_1 + k\right) \\%, \;\;   &r=1,\hdots,b_1 \\
 \theta_{b_1+1}, \theta_{b_1+2},\ldots, \theta_{b_2} &= 2  &  \text{where }   x_{b_1+1}, x_{b_1+2},\ldots, x_{b_2}  \in \left[x_1 + k, x_1 + 2k\right) \\%, \;\;  & r=b_1+1,\hdots,b_2  \\
 & \vdots  &  \\
%  \theta_{b_{m-2}+1},  \theta_{b_{m-2}+2},\ldots, \theta_{b_{m-1}}&= m-1 &  \text{where } x_1 + (m-2)k <  x_r \leq x_1 + (m-1)k, r=b_{m-2}+1,\hdots,b_{m-1} \\
 \theta_{b_{m-1}+1},  \theta_{b_{m-1}+2},\ldots, \theta_{b_{m}} & = m  & \text{where } x_{b_{m-1}+1},  x_{b_{m-1}+2},\ldots, x_{b_{m}}  \in \left[ x_1 + (m-1)k,x_n\right]. \\%, \;\;  & r=b_{m-1}+1,\hdots,b_{m}=n.
\end{eqnarray*}

When the algorithm considers each point $x_i$ for the minimum $f$ in the window spanning $k/2$ either side, it checks whether the most extreme $x$-values in this window are either in the current block or one block away (these values cannot be further than one block away as block sizes are of length $k$) to decide on which combination of $g$ and $h$ is required. 
Note the algorithm is impervious to arbitrarily spaced $x_i$ as long as they are in ascending order. 
If $\theta_i \neq j$ for any $i=2,3,\hdots,n-1; j=2,3,\ldots,m-1$ (empty blocks) or $x_{b_{j-1}+1}=x_{b_{j}}$ for any $j=2,3,\ldots,m$ (blocks with only one $x_i$), for example, do not affect the validity of the proposed algorithm. 

This algorithm can be seen as a generalised version of the LSA \cite{vanHerk1992,Gil1993} as it works on evenly and unevenly spaced data. 
An \texttt{R} implementation of this novel CLSA can be found as an \texttt{R}-package using compiled \texttt{C} code at \texttt{https://github.com/tystan/clsa}.

A demonstration of why the creation of blocks the size of the SE and accessing cumulative values half an SE length away allows the calculation of rolling minimums is shown in \cite{vanHerk1992}.
Examples to demonstrate the mechanics of the CLSA algorithm are presented in Appendix~\ref{appb}.

\FloatBarrier

\section*{Results}

Presented in this paper is a pipeline to automate the baseline subtraction step in proteomic TOF-MS pre-processing. 
The pipeline consists of transforming the {\mz}-axis, then finding an appropriate SE size via an automated peak width estimation algorithm on the transformed scale, applying a novel algorithm to perform the top-hat baseline subtraction, then finally, baseline subtracted spectra are returned by back-transforming the data to the {\mz} scale.

There remain two elements of the pipeline to be assessed. Firstly, for the pipeline to be fully automated, an optimal combination of EPCP value and transformation need to be found. In the next section we perform a grid search over EPCP values of $0.8,0.85,0.9,0.95,0.98,0.99,1$ and transformations $t_0,t_1,t_2,t_3,t_4,t_5$ to find which combination provides the closest baseline subtracted signal to the gold-standard.
Given sufficient similarity to the gold-standard is achieved, it is hoped that a consensus over all datasets, in their varying attributes, of the optimal combination of EPCP value and transformation can be found. 
If a consensus is indeed found, the pipeline is likely to be applicable to other proteomic TOF-MS datasets.

A theoretical and empirical assessment of the efficiency of the CLSA in comparison to naive rolling window algorithms then follows. The theoretical efficiency is discussed with respect to the number of operations required over all the elements input into the CLSA. By performing the top-hat operation on the six proteomic TOF-MS datasets and simulated datasets of varying sizes, the computational time required for the CLSA versus the naive algorithm provides an empirical assessment of their relative efficiencies. 

\subsection*{Comparison of piecewise and transformed axis baseline subtraction}

\begin{figure}[hp] \centering
\includegraphics[width=0.8\textwidth]{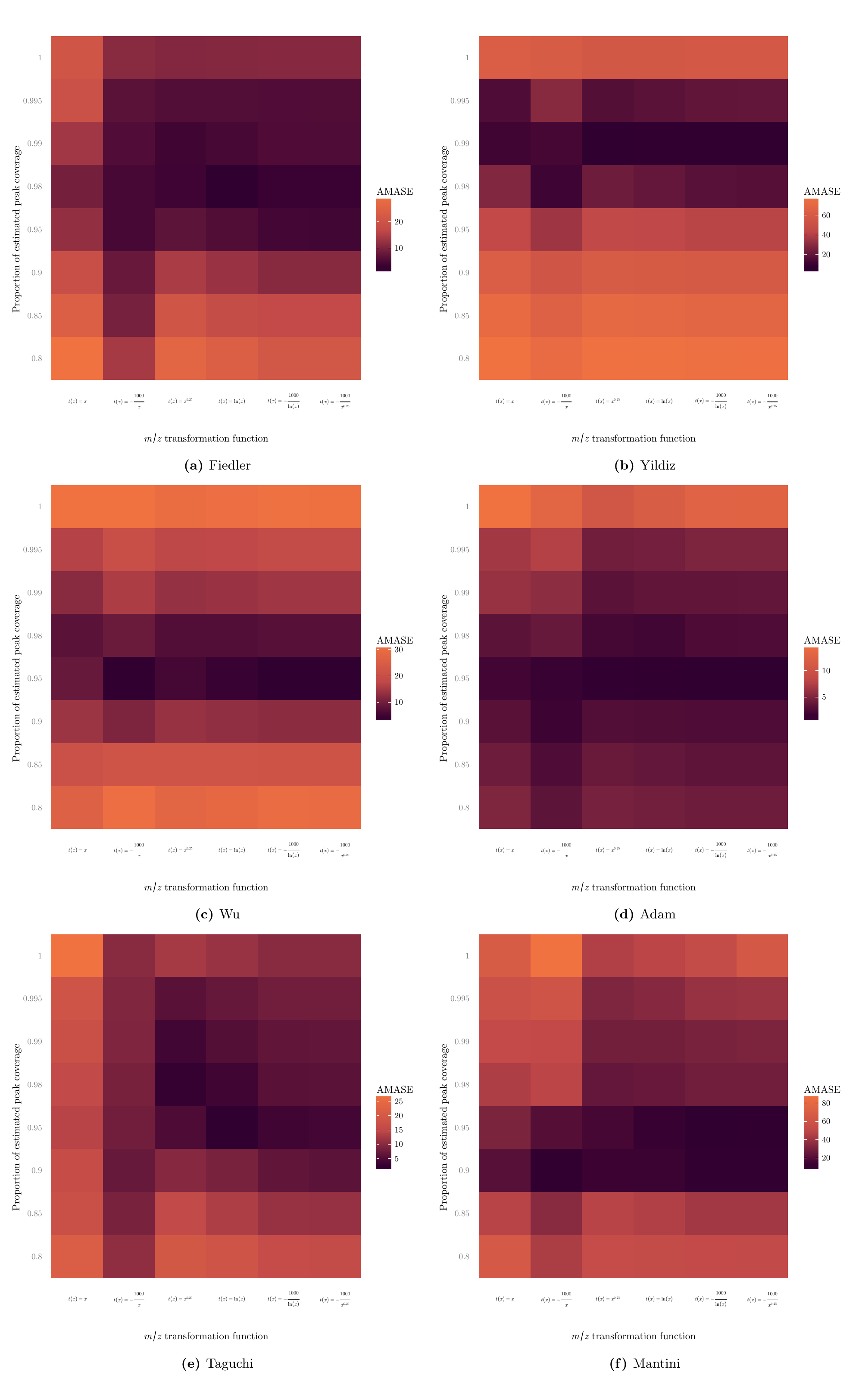}
    \caption{\textbf{Average mean absolute scaled error (AMASE) heatmaps}:
     AMASE values for each of the six datasets under different combinations of {\mz} transformation and estimated peak coverage proportion (EPCP).} \label{masefig}
\end{figure}

Figure~\ref{masefig} and Table~\ref{masetab} present the AMASE values on the six datasets. 
For each dataset, Figure~\ref{masefig} displays a grid of the input arguments which are the {\mz}-axis transformation and the EPCP. 
The colour of the blocks at the intersection of these combinations depict the AMASE value obtained.
Darker blocks indicate smaller, and thus preferred, AMASE values. Table~\ref{masetab} is simply a tabular presentation of the results shown in Figure~\ref{masefig}.

\begin{table}[h!] 
\centering \scriptsize
\caption{Average mean absolute scaled error (AMASE) when using structuring element (SE) sizes corresponding to different estimated peak coverage proportions (EPCPs) for each of the selected short-listed transformations on each of the six datasets.} \label{masetab}
\begin{tabular}{rrrrrrr}
  \hline
EPCP & $t_0\left(x\right)=x$ & $t_1\left(x\right)=\frac{-1000}{x}$ & $t_2\left(x\right)=x^{1/4}$ & $t_3\left(x\right)=\ln x$ & $t_4\left(x\right)=\frac{-1000}{\ln x}$ & $t_5\left(x\right)=\frac{-1000}{x^{1/4}}$  \\ 
\hline
\multicolumn{7}{c}{Fiedler}\\
  \hline
1 & 19.8 & 10.1 & 9.5 & 9.7 & 9.9 & 9.9 \\ 
  0.995 & 18.4 & 5.5 & 4.8 & 4.7 & 4.7 & 4.6 \\ 
  0.99 & 12.4 & 4.7 & 3.0 & 3.7 & 4.5 & 4.5 \\ 
  0.98 & 8.2 & 3.7 & 2.8 & 1.3 & 2.3 & 2.3 \\ 
  0.95 & 11.0 & 3.7 & 5.8 & 4.6 & 3.3 & 3.1 \\ 
  0.9 & 17.8 & 6.9 & 13.2 & 11.7 & 9.9 & 10.0 \\ 
  0.85 & 23.2 & 8.3 & 19.8 & 16.9 & 16.1 & 16.2 \\ 
  0.8 & 29.4 & 12.9 & 25.5 & 22.9 & 20.2 & 20.2 \\ 
   \hline
\multicolumn{7}{c}{Yildiz}\\
  \hline   
1 & 59.7 & 58.3 & 54.9 & 54.6 & 55.6 & 55.7 \\ 
  0.995 & 12.9 & 27.4 & 14.3 & 15.8 & 17.5 & 18.0 \\ 
  0.99 & 9.3 & 10.7 & 5.0 & 5.2 & 5.5 & 5.5 \\ 
  0.98 & 25.7 & 8.6 & 20.9 & 18.4 & 15.4 & 14.9 \\ 
  0.95 & 42.3 & 32.6 & 41.6 & 41.0 & 39.7 & 39.5 \\ 
  0.9 & 59.7 & 52.7 & 58.4 & 57.7 & 56.7 & 56.6 \\ 
  0.85 & 70.4 & 62.9 & 69.2 & 68.1 & 67.2 & 67.0 \\ 
  0.8 & 77.1 & 71.5 & 76.6 & 75.9 & 75.0 & 74.9 \\ 
   \hline
\multicolumn{7}{c}{Wu}\\
  \hline  
1 & 30.9 & 30.8 & 29.2 & 29.7 & 30.4 & 30.2 \\ 
  0.995 & 16.1 & 19.5 & 16.9 & 17.2 & 18.0 & 17.9 \\ 
  0.99 & 12.0 & 15.2 & 13.0 & 13.5 & 14.1 & 14.0 \\ 
  0.98 & 7.4 & 9.0 & 6.7 & 6.7 & 7.2 & 7.1 \\ 
  0.95 & 8.8 & 3.4 & 5.4 & 4.1 & 3.3 & 3.3 \\ 
  0.9 & 13.8 & 10.9 & 13.3 & 12.8 & 12.3 & 12.3 \\ 
  0.85 & 20.1 & 20.9 & 20.9 & 21.1 & 20.7 & 20.7 \\ 
  0.8 & 25.4 & 29.7 & 27.0 & 27.7 & 28.8 & 28.7 \\ 
   \hline
\multicolumn{7}{c}{Adam}\\
  \hline 
1 & 14.8 & 12.7 & 10.1 & 11.2 & 12.3 & 12.4 \\ 
  0.995 & 6.2 & 7.1 & 3.8 & 3.9 & 4.5 & 4.5 \\ 
  0.99 & 5.7 & 5.2 & 2.7 & 3.0 & 3.1 & 3.1 \\ 
  0.98 & 2.8 & 3.4 & 1.6 & 1.5 & 2.1 & 2.1 \\ 
  0.95 & 1.5 & 1.0 & 0.7 & 0.6 & 0.6 & 0.5 \\ 
  0.9 & 2.6 & 1.2 & 2.3 & 2.2 & 2.1 & 2.1 \\ 
  0.85 & 3.7 & 2.2 & 3.4 & 3.2 & 2.9 & 2.9 \\ 
  0.8 & 4.6 & 2.9 & 4.1 & 3.9 & 3.7 & 3.7 \\ 
   \hline
\multicolumn{7}{c}{Taguchi}\\
  \hline 
1 & 26.5 & 9.6 & 12.0 & 11.0 & 9.6 & 9.7 \\ 
  0.995 & 17.7 & 8.9 & 5.5 & 6.5 & 7.5 & 7.5 \\ 
  0.99 & 16.4 & 8.8 & 3.5 & 5.1 & 6.1 & 6.2 \\ 
  0.98 & 14.8 & 8.1 & 2.2 & 3.3 & 5.5 & 5.6 \\ 
  0.95 & 13.6 & 7.6 & 4.5 & 1.8 & 3.3 & 3.6 \\ 
  0.9 & 15.7 & 6.7 & 9.4 & 8.2 & 6.2 & 5.8 \\ 
  0.85 & 16.5 & 8.2 & 14.7 & 12.7 & 10.8 & 10.7 \\ 
  0.8 & 20.7 & 10.2 & 18.9 & 17.7 & 15.4 & 14.9 \\ 
   \hline
\multicolumn{7}{c}{Mantini} \\
  \hline  
1 & 66.1 & 86.4 & 42.8 & 45.6 & 50.6 & 63.0 \\ 
  0.995 & 55.1 & 58.2 & 29.6 & 31.8 & 36.3 & 37.2 \\ 
  0.99 & 49.3 & 48.3 & 26.2 & 26.2 & 27.9 & 28.9 \\ 
  0.98 & 42.1 & 45.9 & 22.3 & 23.4 & 25.8 & 25.9 \\ 
  0.95 & 28.8 & 18.1 & 13.9 & 9.5 & 7.6 & 7.9 \\ 
  0.9 & 18.8 & 8.1 & 11.0 & 10.6 & 8.5 & 8.5 \\ 
  0.85 & 45.3 & 32.4 & 45.0 & 42.9 & 38.9 & 38.9 \\ 
  0.8 & 63.1 & 41.6 & 51.2 & 49.7 & 49.2 & 49.0 \\ 
   \hline
\end{tabular}
\end{table}

No single transformation or EPCP was optimal.
However, EPCP between 0.95 and 0.99 provided the optimal AMASE value for all datasets suggesting the peak width estimation process is relatively stable.
On the Fiedler, Yildiz, Taguchi and Mantini datasets, the null transformation which implicitly implies a constant peak width across the {\mz}-axis is not valid as AMASE values are notably higher than for the remaining transformations.
The transformations $t_2$, $t_3$, $t_4$ and $t_5$ produced the best results.
It should be noted that the transformations $t_3$, $t_4$ and $t_5$ produced very similar AMASE values.
With the exception of the Yildiz dataset, using these transformations with an EPCP of 0.95 produced sensible results.

\begin{figure}[h!] \centering
 \includegraphics[width=\textwidth]{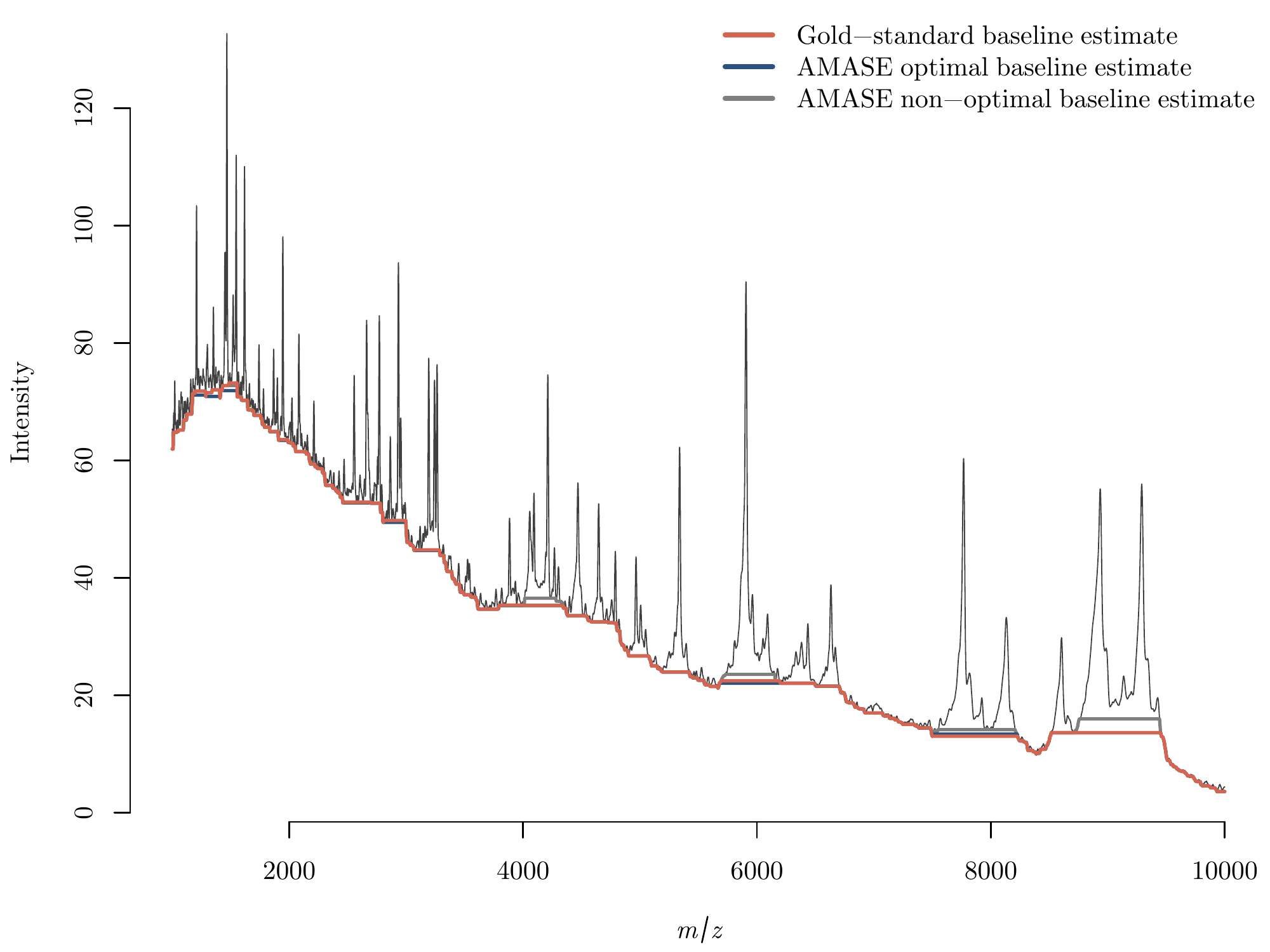}
  \caption{\textbf{Optimised automated baseline estimate (blue) in comparison to the gold-standard (orange) piecewise baseline estimate for the Fiedler spectrum}:
    the optimal transformation and EPCP were $t_3$ and 0.98, respectively; the non-optimal combination (grey) of transformation and EPCP shown is $t_4$ and 0.95, respectively.} \label{fig:amasevgs} 
\end{figure}

Figure~\ref{fig:amasevgs} demonstrates the baseline estimates using the gold-standard piecewise top-hat operator, the AMASE optimal transformation and EPCP ($t_3$, 0.98) and a non-optimal combination of transformation and EPCP ($t_4$, 0.95) that was suitable on all but the Yildiz data.
The optimal AMASE transformation and EPCP combination ($t_3$, 0.98) shows very little difference from the gold-standard baseline estimate.

Because the gold-standard baseline estimate is subject to expert input and opinion, the differences seen in the gold-standard and the optimal AMASE baseline estimate are not of concern as both look sensible. %negligible when observed
The non-optimal baseline estimate produces a reasonable automated baseline subtraction, however, it can be seen that this estimate does undercut the peaks especially at high {\mz}-values.

With respect to the AMASE values, the spectra with fewer peaks generally had larger AMASE values; this is a function of the normalising constant for each spectra, $\frac{1}{n-1} \sum_{i=2}^n  | \tau^{*}_i- \tau^{*}_{i-1} |_{j=1,2,\hdots,n}$, as fewer peaks will generally imply less relative change in signal.

\FloatBarrier

\subsection*{Efficiency of the CLSA compared to the naive rolling window}

The naive rolling minimum algorithm consists of the linear-time process of finding the indexes of points at the upper and lower edges of the sliding window for each element, by incrementing the edge indexes from the previous element when required.
Using $a_k$ as the average number of data points in the sliding window of size $k$, the computational cost of finding the minimum value in the window requires approximately $a_k-1$ comparisons per element. 
This is because each element requires, on average, a minimum or maximum comparison of all the data points in the window except one: the first data point does not require a comparison. 
The resulting computational complexity is $\mathcal{O}\left(a_kn\right)$ for the naive algorithm, which is dependent on the size of the sliding window and the number of elements in $X$.

Like the LSA, the CLSA is a linear-time algorithm irrespective of the window size, $k$. 
For the CLSA, a linear-time progression through the $n$ elements is required to assign integers of the $\Theta$-vector, as each element is an integer equal to or greater than that which precedes it. 
The linear-time process of finding the $W^\il$ and $W^\ir$ indexes at the lower and upper edges of the sliding window, respectively, for each element is similar to that required in the naive algorithm. 
One linear-time sweep forward and one linear-time sweep back on the data is required to create $g$ and $h$. 
A final sweep of the created vectors $W^\il$, $W^\ir$, $\Theta$, $g$ and $h$ is required to compute the $r_\text{min}$ values. 
Each $r_\text{min}\left(f\left(x_i\right)\right)$ calculation requires the tests $\theta_{w^\il_i} = \theta_{w^\ir_i+1}$, $\theta_{w^\il_i-1} = \theta_{w^\ir_i}$ or $\min\left\{g\left(x_i\right),h\left(x_i\right)\right\}$. 
It can therefore be deduced the CLSA is $\mathcal{O}(n)$ complexity, requiring a series of linear-time operations, importantly independent of the length of the sliding window, $k$. 

Given the MS application, $a_k-1$ operations per element in the naive algorithm would be much larger than the constant number of operations required per element for the CLSA and efficiency strongly favours the CLSA. 
It should be pointed out that the CLSA requires extra memory availability beyond the iterative algorithm for the creation of the vectors $W^\il$, $W^\ir$, $\Theta$, $g$ and $h$. 
Another computational advantage of the CLSA is that by using the minimum of the two temporary vectors $g$ and $h$ as opposed to the minimum of a non-constant number of data points for each $x_i \in X$, vectorised programming can be utilised instead of loops. 
This is of significant advantage in programming languages that are interpreted such as \texttt{R}. 

Using the \texttt{clsa} package, the CLSA and naive sliding window algorithms were compared for computational time to calculate the top-hat on real and simulated data. 
The computations were performed on a 21.5'' iMac (late 2013 model, 2.7GHz Intel Core i5, 8GB 1600MHz DDR3 memory, OS X 10.10.2). 
To optimise speed, the calculations requiring iterative looping were performed using compiled \texttt{C} code for both the CLSA and naive algorithms. 
The code to run the test of computational running time on the simulated data is provided in Appendix~\ref{appc}.

The CLSA and naive sliding window algorithms were applied to perform top-hat baseline estimation to the six datasets used in this paper and  
the results are shown in Table~\ref{tab:clsa}. 
The CLSA resulted in a reduction of the required computational time by a factor of at least  4. 
The advantage in speed of the CLSA had greater improvement for the datasets with a greater number of {\mz} values.
The biggest relative improvement was by a factor exceeding 50 for the largest dataset in terms of {\mz} values per spectra on the Yildiz spectra.

\begin{table}[h!]\centering
\caption{Computational time to perform top-hat baseline subtraction in the transformed space using the naive and CLSA algorithms on the six datasets under study.} 
\label{tab:clsa}
      \begin{tabular}{lrrrr}
        \hline
          & Number of & Number of  & \multicolumn{2}{c}{Computational time (sec)}  \\ \cline{4-5}
         Data & specta & {\mz} values & Naive algorithm  & CLSA  \\ \hline
        Fiedler  &  16  &  42388  &   7.7 & 0.2  \\ 
        Yildiz  & 264  &  75958  &  312.6  &   5.5  \\   
        Wu  & 89  &  91378  & 34.1  & 1.7   \\ 
        Adam  & 326  &  8461  & 3.0 &  0.7    \\
        Taguchi &  210  &  19234  & 18.0 &  0.9   \\  
        Mantini  & 30  &  32967  & 6.7 &  0.2     \\
        \hline
      \end{tabular}
\end{table}

Table~\ref{tab:clsa:synth} displays the computational times of top-hat baseline estimation using the CLSA and naive algorithms for varying datasets and  SE sizes. 
The simulated data consisted of 20 randomly generated spectra with $x_i$ and $f_i$ for $i=1,2,\hdots,n$. 
These values were independently and randomly generated, where the signal locations $x_i \sim \text{Beta}\left(1,3\right)$  mimic a higher density of points at the low end of the spectra and $f_i \sim \chi^2_{10}$  mimic the positive signals in spectra. 
MALDI TOF-MS data can have in excess of tens of thousands of {\mz} values, hence, values of $n=10^4,2\times{}10^4,\hdots,10^5$ were used. Varying window sizes were tested, ranging in width from 0.5\% to 20\% of the $x$-axis domain. i.e.,  0.5\% corresponds to a window size of $0.005$ passed over the domain $\left[0,1\right]$.

\begin{table}[h!]
\centering
\caption{Computational time in seconds to perform top-hat baseline subtraction in the transformed space using the  naive and CLSA algorithms on  synthetic data for varying data assumptions and SE sizes.}
\label{tab:clsa:synth}
{\footnotesize
\begin{tabular}{l|rrrrrrcrrrrrr}
  \hline
Number of   & \multicolumn{6}{c}{Naive} &  & \multicolumn{6}{c}{CLSA} \\
points & \multicolumn{6}{c}{Window size (\% of $x$-axis)} & & \multicolumn{6}{c}{Window size (\% of $x$-axis)} \\
 $n$ ($\times 10^4$)  & 0.5 & 1 & 2 & 5 & 10 & 20 &  & 0.5 & 1 & 2 & 5 & 10 & 20 \\ 
  \hline
1 & 0.1 & 0.1 & 0.3 & 0.6 & 1.2 & 2.3 &  & 0.0 & 0.0 & 0.2 & 0.0 & 0.0 & 0.2 \\ 
  2 & 0.3 & 0.5 & 1.0 & 2.5 & 4.9 & 9.2 &  & 0.1 & 0.1 & 0.1 & 0.1 & 0.1 & 0.1 \\ 
  3 & 0.6 & 1.2 & 2.3 & 5.7 & 11.0 & 20.6 &  & 0.3 & 0.1 & 0.1 & 0.2 & 0.1 & 0.1 \\ 
  4 & 1.1 & 2.1 & 4.2 & 10.2 & 19.6 & 36.5 &  & 0.1 & 0.2 & 0.1 & 0.1 & 0.3 & 0.1 \\ 
  5 & 1.7 & 3.3 & 6.5 & 15.8 & 30.5 & 56.8 &  & 0.2 & 0.3 & 0.2 & 0.2 & 0.3 & 0.2 \\ 
  6 & 2.4 & 4.7 & 9.3 & 22.7 & 44.0 & 82.0 &  & 0.2 & 0.2 & 0.3 & 0.2 & 0.2 & 0.3 \\ 
  7 & 3.2 & 6.4 & 12.6 & 30.9 & 59.7 & 111.7 &  & 0.2 & 0.2 & 0.4 & 0.2 & 0.2 & 0.4 \\ 
  8 & 4.2 & 8.4 & 16.6 & 40.3 & 78.3 & 146.4 &  & 0.4 & 0.3 & 0.3 & 0.4 & 0.3 & 0.3 \\ 
  9 & 5.4 & 10.6 & 21.0 & 51.1 & 98.8 & 185.3 &  & 0.4 & 0.3 & 0.3 & 0.4 & 0.3 & 0.3 \\ 
  10 & 6.6 & 13.0 & 25.9 & 63.1 & 121.8 & 228.7 &  & 0.3 & 0.4 & 0.3 & 0.3 & 0.5 & 0.3 \\ 
   \hline
\end{tabular}
}
\end{table}

The CLSA was faster than the naive algorithm in every scenario as shown in Table~\ref{tab:clsa:synth}. 
As expected, the computational time was constant for the CLSA irrespective of the window size for a fixed number of points (number of transformed {\mz} values).
The difference in computational time between the two algorithms was reasonably small for small datasets and small SE sizes.
However, for a typical number of {\mz} points seen in practice, say 50,000, and a moderate window size that on average encapsulates 5,000 points (1\% of $x$-axis), the CLSA provides an order of magnitude increase in speed.

\FloatBarrier

\section*{Discussion}

The current gold-standard  in  baseline subtraction is a piecewise approach that is performed manually, that is, by inspection. Piecewise baseline subtraction is typically performed because, as we have consistently observed with the datasets analysed in this paper, the properties of the spectra do not remain constant over their domain. In particular, a spectrum's peak width increases with increasing {\mz}-values. 
We  have proposed  a  new baseline subtraction pipeline be adopted for  the correction of mass proteomic spectra data  which avoids both the manual user input and the piecewise-subtraction aspect of existing methods. Our new pipeline is based on the premise that a suitable transformation of the {\mz}-axis can be found which removes  the relationship between peak width and peak location.

As part of the new pipeline, we propose a method to create data-based, and therefore data specific, peak-width estimates from smoothed spectra. Even if this step is not used to automate  baseline subtraction, it provides an initial sensible SE size that  adapts to each individual dataset. 
Our generalised version of the LSA is also presented in the paper, which we call CLSA.  CLSA can be applied to unevenly or evenly spaced data and is not limited in its application to proteomic MS data.  Should a transformation be known to create peak widths independent of {\mz}-location in proteomic MS data, an efficient and effective baseline subtraction can be performed using the top-hat operator with a CLSA implementation.   A major contribution to note is that we have demonstrated CLSA far outperforms the naive rolling minimum algorithm in required computational time by an order of magnitude or more on numerous   datasets of real-world complexity.

 The transformed and constant-sized window approach may suffer from a slight but largely unnoticeable reduction in sensitivity in comparison. 
The trade-off between exactness of the piecewise approach and the speed of the automated transformation and continuous approach may be a consideration, especially if a known relationship exists between the peak width and peak location.

\section*{Availability of supporting data}

\begin{description}
\item[Fiedler data:] A subset of the MALDI TOF-MS data generated by the study~\cite{Fiedler2007} is available in the publicly available \texttt{R} package: \texttt{MALDIquant}~\cite{MALDIquant}.
\item[Yildiz data:] Available at \\ {\webtext{http://www.vicc.org/biostatistics/serum/JTO2007.htm}}.
\item[Wu data:] Previously available at \\ {\webtext{http://bioinformatics.med.yale.edu/MSDATA}}.
\item[Adam data:] Data was obtained on request from the authors of \cite{Adam2002}. However some Eastern Virginia Medical School data is available at \\ 
{\webtext{http://edrn.nci.nih.gov/science-data}}.
\item[Taguchi data:] Available at  \\
{\webtext{http://www.vicc.org/biostatistics/download/WSData.zip}}.
\item[Mantini data:] Available at \\ {\webtext{http://www.biomedcentral.com/content/supplementary/1471-2105-8-101-S2.zip}}.
\end{description}

\section*{Competing interests}
  The authors declare that they have no competing interests.

\section*{Author's contributions}

TS and PS developed the statistical and analytical methods. CB and PS provided guidance on the analysis of proteomic data. TS developed the code and implementation. All authors contributed to the writing of the manuscript.

\section*{Acknowledgements}

Thank you to the creators and custodians of the publicly available data used in this manuscript.
TS's PhD research was funded by an Australian Postgraduate Award scholarship.

%%%%%%%%%%%%%%%%%%%%%%%%%%%%%%%%%%%%%%%%%%%%%%%%%%%%%%%%%%%%%%%%%%%%%%%%%%%%%%%%%%%%%%%%%%%%%%%%

\bibliography{StanfordBagleySolomon2015}

%%%%%%%%%%%%%%%%%%%%%%%%%%%%%%%%%%%%%%%%%%%%%%%%%%%%%%%%%%%%%%%%%%%%%%%%%%%%%%%%%%%%%%%%%%%%%%%%

\clearpage

% Set up the naming conventions for equations in the appendices
\setcounter{table}{0}
\setcounter{figure}{0}
\renewcommand{\theequation}{\Alph{section}.\arabic{equation}} %\arabic{section}.
\renewcommand\thetable{\thesection.\arabic{table}}    
\renewcommand\thefigure{\thesection.\arabic{figure}}

\begin{appendices}

\section{Six transformation functions on each of the six datasets} \label{appa}

The effect of the six transformation functions, $t_0$-$t_5$, on a spectrum from each of the six datasets are shown in Figures~\ref{fig12}~to~\ref{fig17}.

\newgeometry{textwidth=20cm,textheight=25cm}

\clearpage

\newcommand{\fighere}[2]{%
\begin{figure}[h!]
  \centering
    \includegraphics[height=\textheight]{fig/fig#1.pdf}
  \caption{The effect of the six transformation functions, $t_0$-$t_5$, on a spectrum from the #2 dataset.} \label{fig#1}
\end{figure}%
}

\pagestyle{empty}

%%%%%%%%%%%%%%%%%%%%% FIGURE %%%%%%%%%%%%%%%%%%%%%%

\fighere{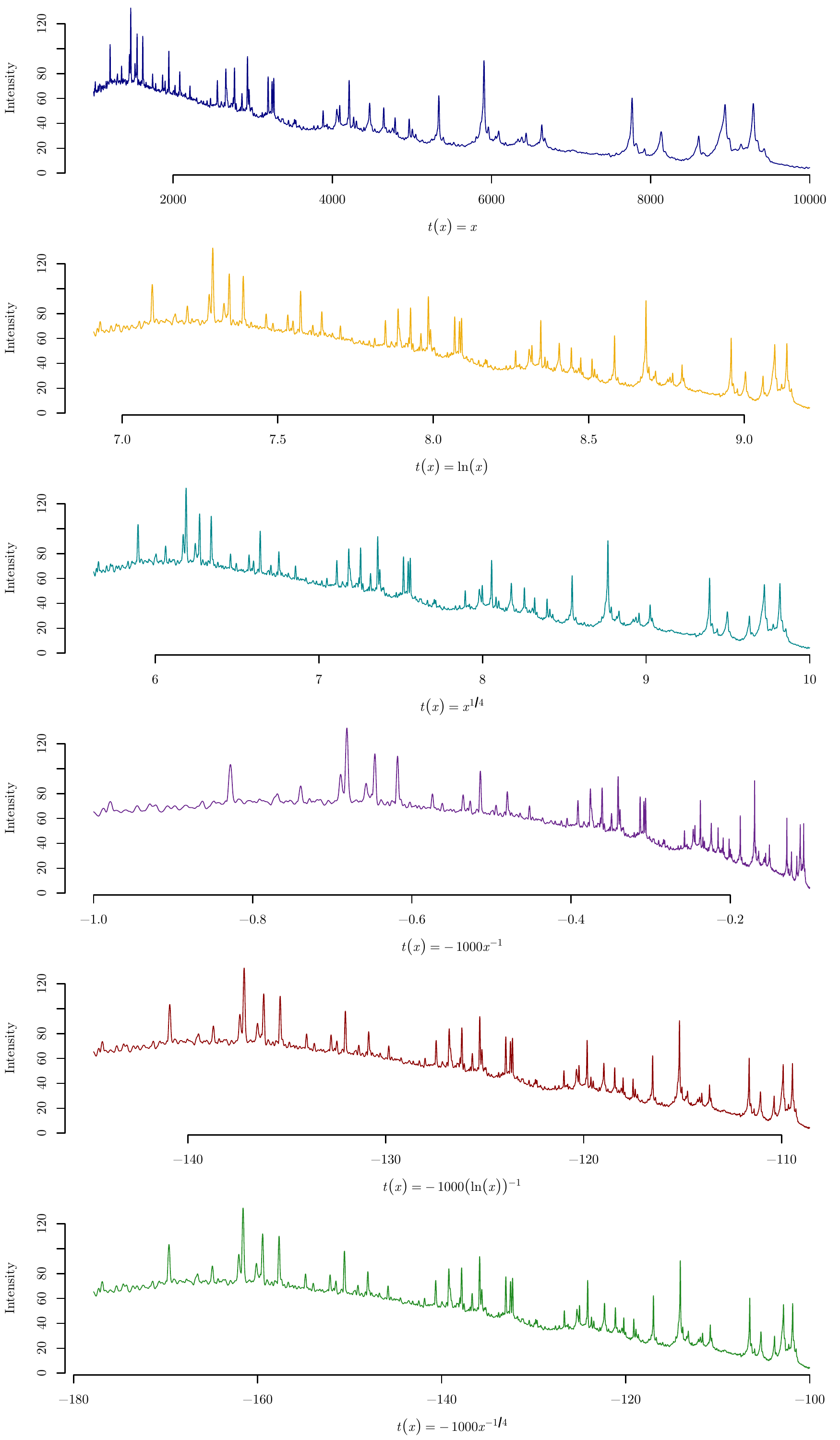}{Fiedler}
\fighere{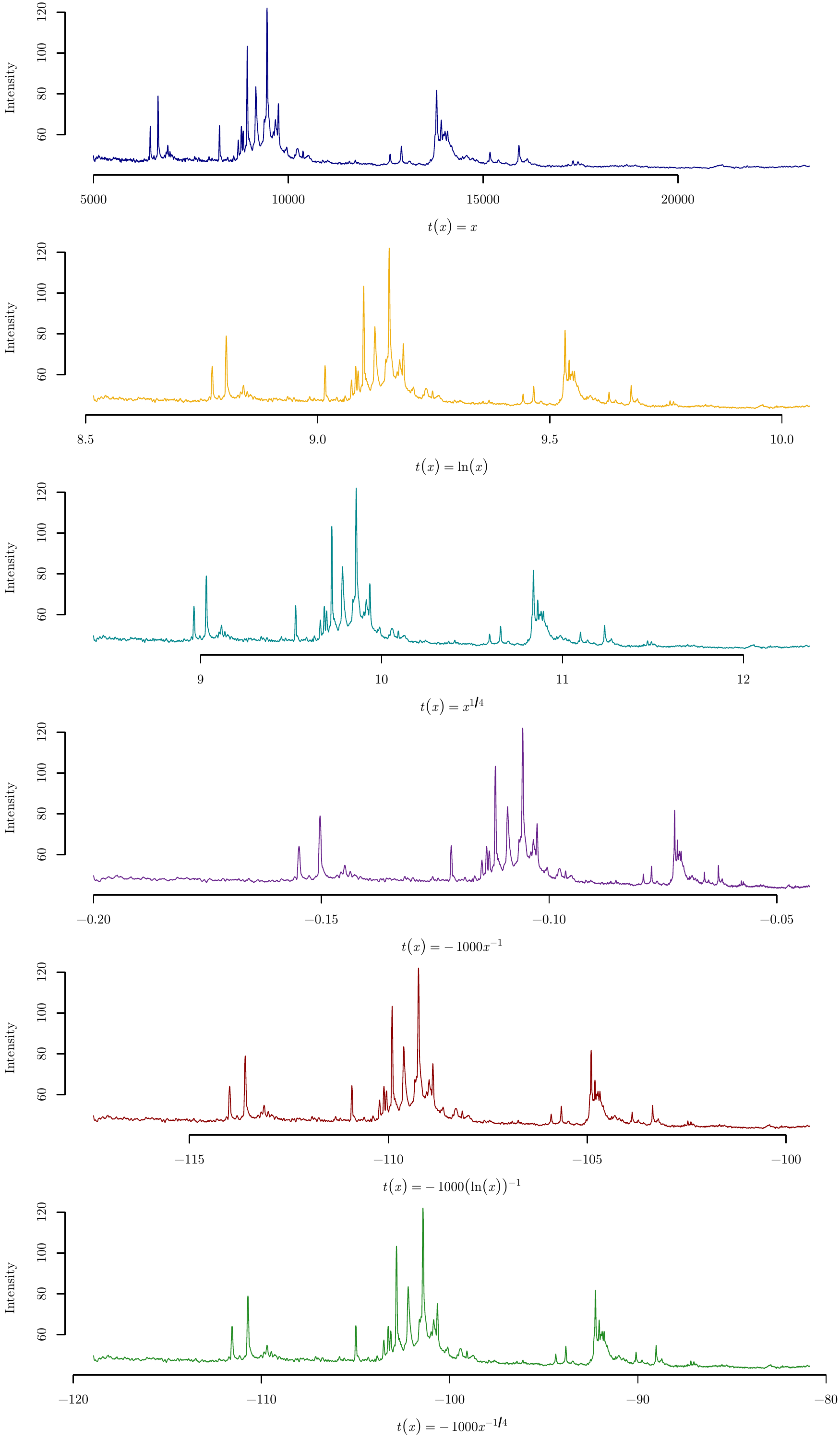}{Yildiz}
\fighere{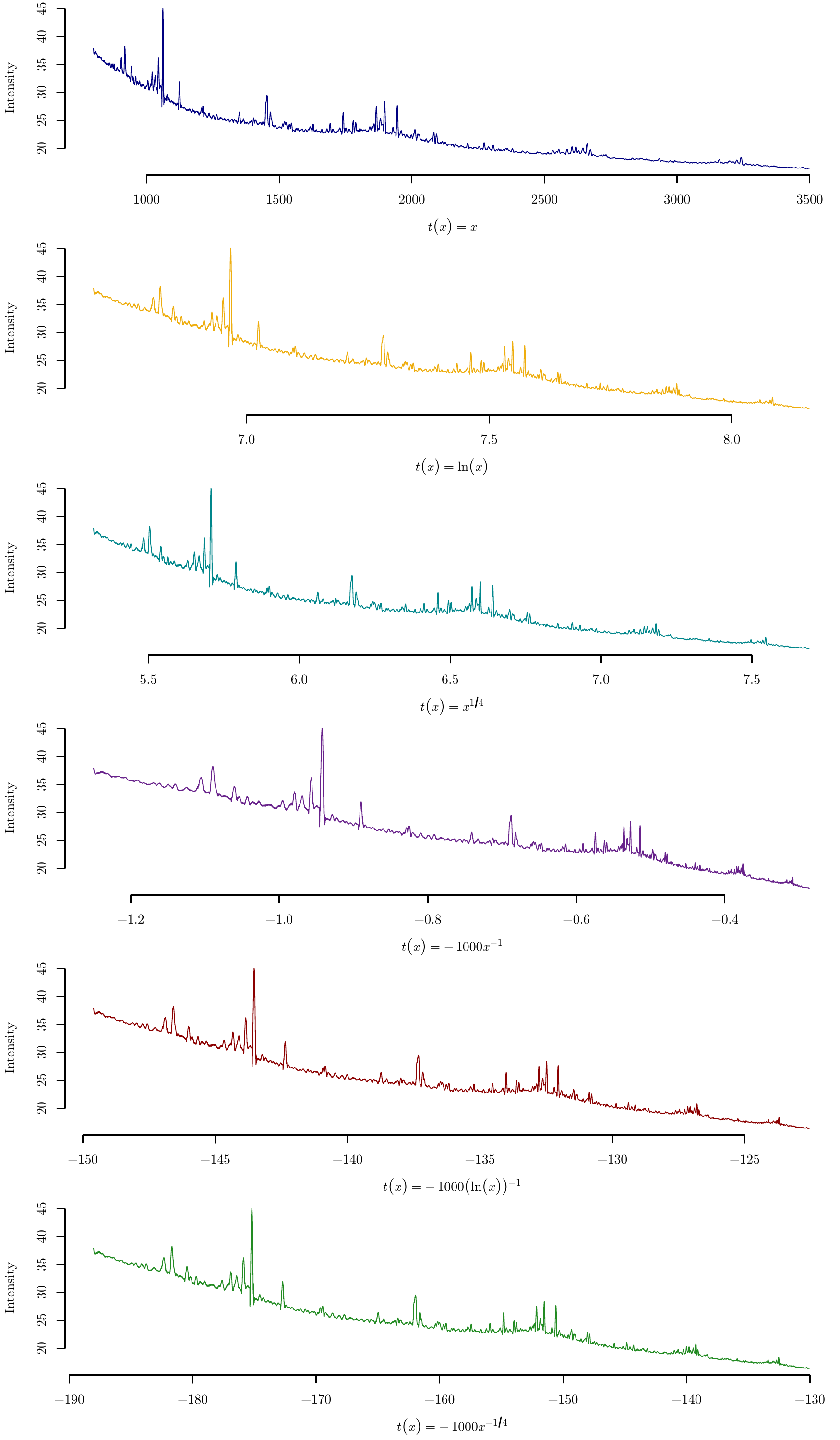}{Wu}
\fighere{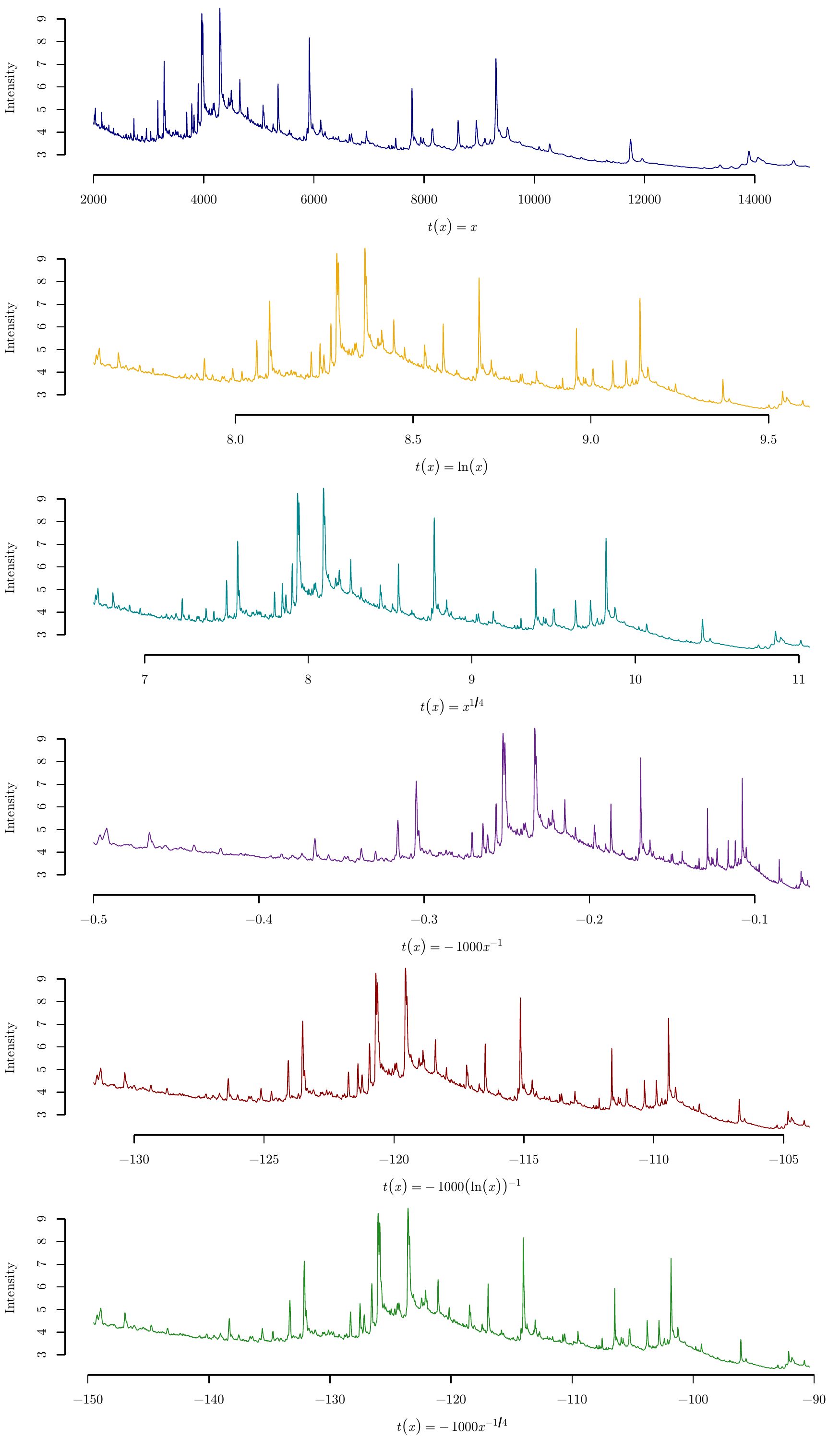}{Adam}
\fighere{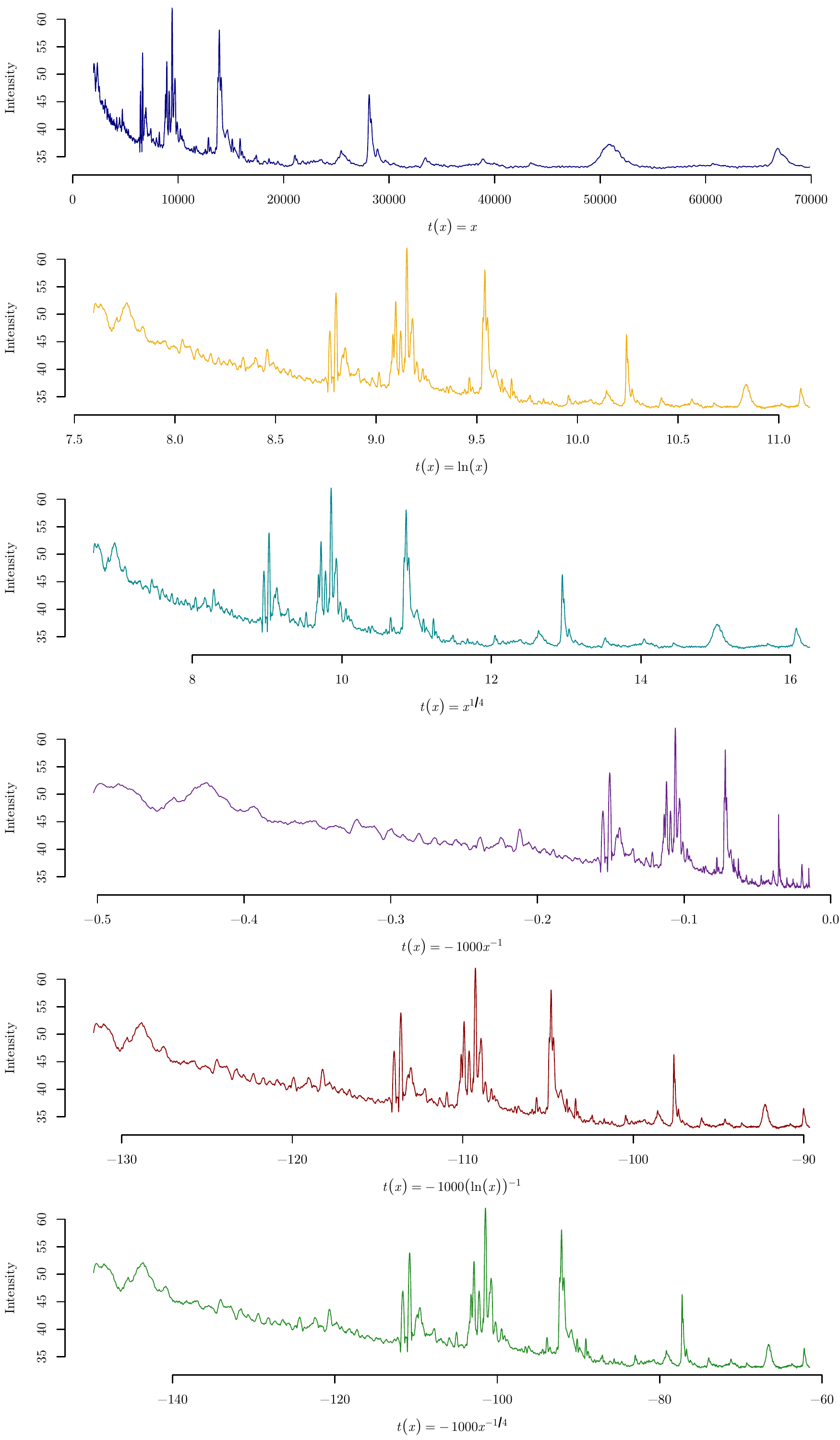}{Taguchi}
\fighere{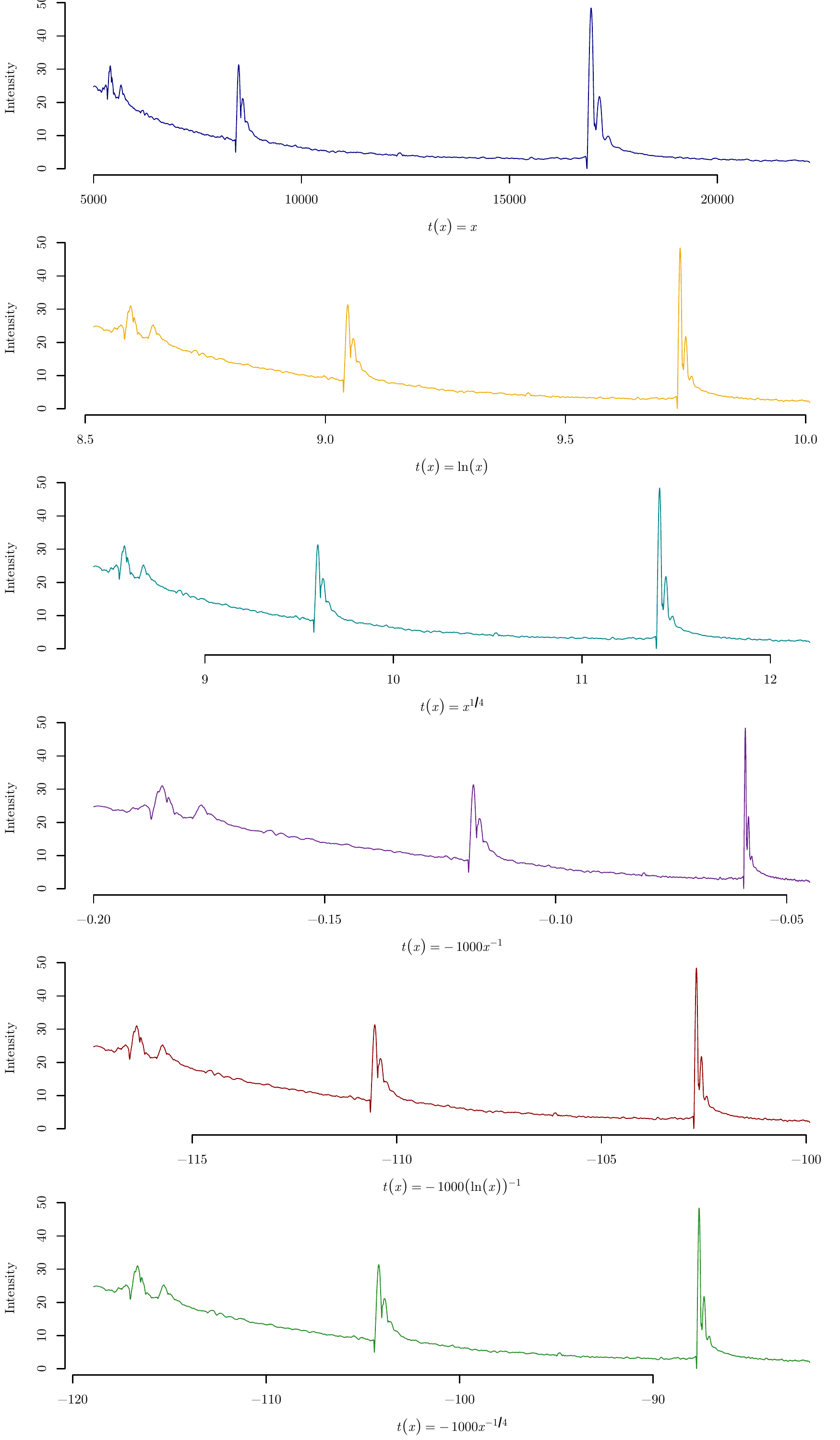}{Mantini}

\restoregeometry
\pagestyle{plain}
\clearpage
\setcounter{table}{0}
\setcounter{figure}{0}

\section{CLSA examples} \label{appb}

\begin{figure}[h!]
  \centering
\includegraphics[width=0.7\textwidth]{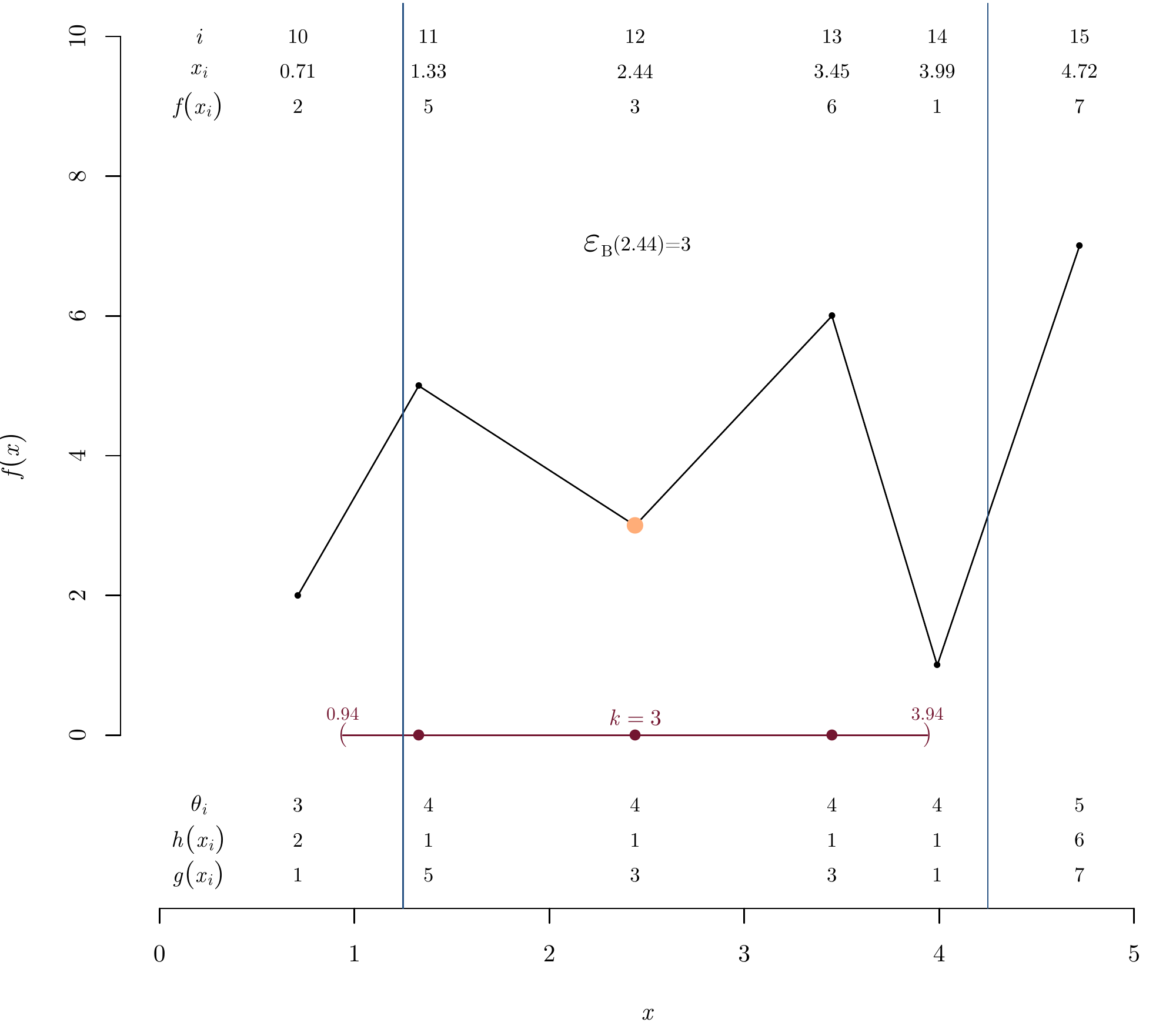}
\caption{An example of data for $x_i=x_{12}=2.44$ and $k=3$ where $\theta_{w_{i}^{\il}} = \theta_{w_{i}^{\ir}+1}$ (i.e. $\theta_{w^{\il}_{12}} = \theta_{w^{\ir}_{12}+1} = 4$) and the computation required to return the result of the CLSA (the tan coloured point $f(2.44)=3$).}
\label{figcts_case2}
\end{figure}

\begin{figure}[h!]
  \centering
\includegraphics[width=0.75\textwidth]{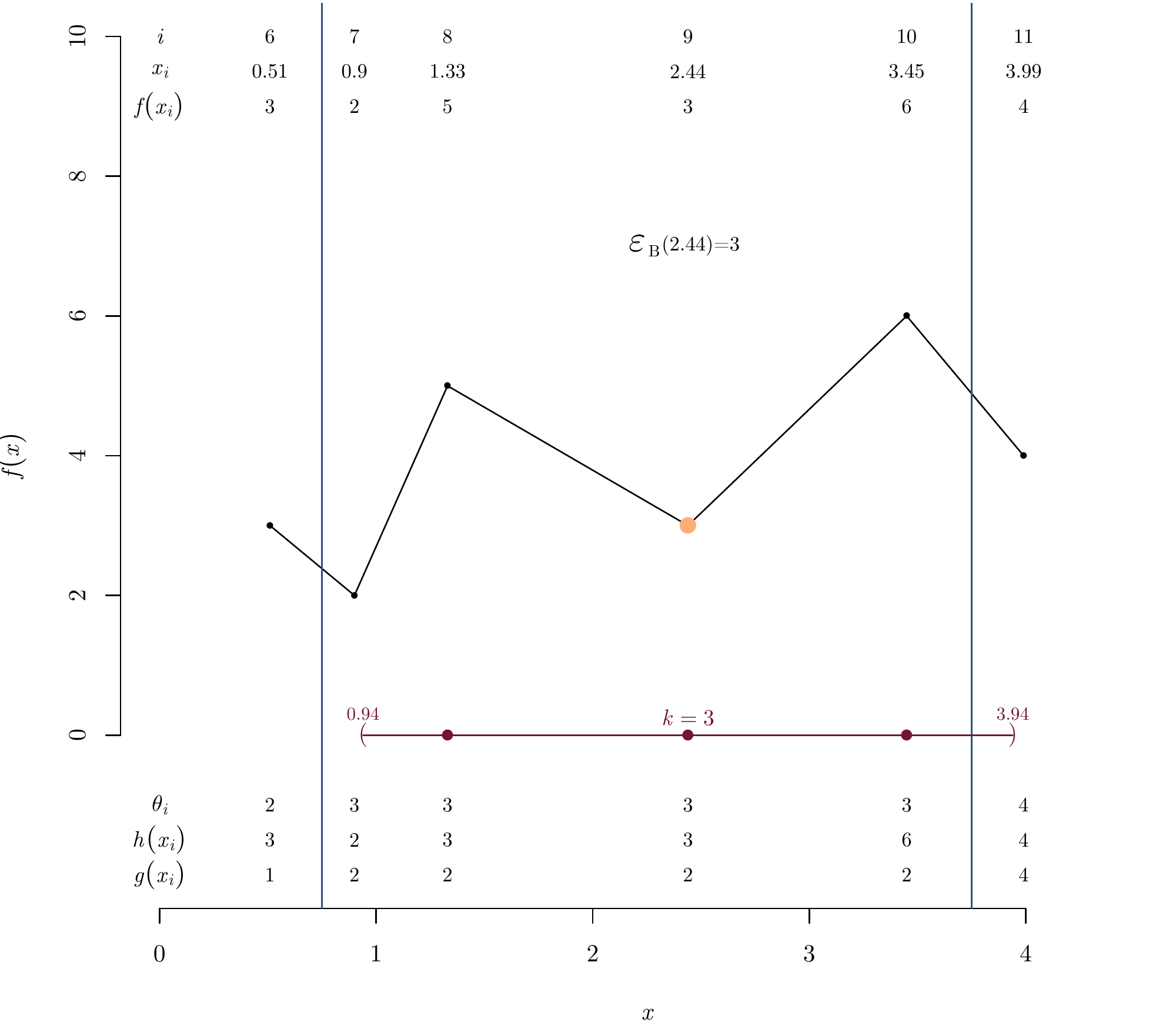}
\caption{An example of data where $\theta_{w_{i}^{\il}-1} = \theta_{w_{i}^{\ir}}$ and the computation required for the continuous line segment algorithm.}
\label{figcts_case1}
\end{figure}

To illustrate how the CLSA works, consider two cases of the algorithm in returning the erosion in Figures \ref{figcts_case2} and \ref{figcts_case1}.

Figure~\ref{figcts_case2} shows a case where $\epsilon_B (f)(x_{12})=3$ using a SE of size $k=3$. It can be seen that,
$$
\theta_{w^{\il}_{12}-1}=\theta_{10}=3 \neq \theta_{w^{\ir}_{12}}=\theta_{13}=4,
$$ 
but,
$$
\theta_{w^{\il}_{12}}=\theta_{11}=3 = \theta_{w^{\ir}_{12}+1}=\theta_{14}.
$$
Therefore the desired result is also achieved using the CLSA as, $$r_{\text{min}} \left( f  \left( x_{12} \right)\right) = g( x_{w^{\ir}_{12}}) = g(x_{13}) = 3 = \epsilon_B (f)(x_{12}).$$

\clearpage

Figure~\ref{figcts_case1} is a different case in the CLSA where $\theta_{w_{i}^{\il}-1} = \theta_{w_{i}^{\ir}}$, as opposed to the case shown in Figure~\ref{figcts_case2} where $\theta_{w_{i}^{\il}} = \theta_{w_{i}^{\ir}+1}$. To obtain the erosion of point $x_i=x_{9}=2.44$ for $k=3$ using the CLSA, observe that 
$$
\theta_{w^{\il}_{9}}=\theta_{8}=3 \neq \theta_{w^{\ir}_{9}+1}=\theta_{11}=4,
$$ 
and 
$$
\theta_{w^{\il}_{9}-1}=\theta_{7}=3 = \theta_{w^{\ir}_{9}}=\theta_{10}.
$$
Therefore, the result of the CLSA erosion is
 $$r_{\text{min}} \left( f  \left( x_{9} \right)\right) = h( x_{w^{\il}_{9}}) = h(x_{8}) = 3.$$

\section{CLSA and naive algorithm computational times for simulated data} \label{appc}

   The \texttt{R} code required to produce the top-hat baseline subtraction computational time results shown in Table~\ref{tab:clsa:synth} is presented below.

\input{app-c.tex}

\end{appendices}

\end{document}

%% file: app-c.tex
\begin{Verbatim}[commandchars=\\\{\},codes={\catcode`\$=3\catcode`\^=7\catcode`\_=8},gobble=0,fontsize=\footnotesize,tabsize=2]
\PY{c+c1}{\PYZsh{}\PYZsh{}\PYZsh{}\PYZsh{}\PYZsh{}\PYZsh{}\PYZsh{}\PYZsh{}\PYZsh{}\PYZsh{}\PYZsh{}\PYZsh{}\PYZsh{}\PYZsh{}\PYZsh{}\PYZsh{}\PYZsh{}\PYZsh{}\PYZsh{}\PYZsh{}\PYZsh{}\PYZsh{}\PYZsh{}\PYZsh{}\PYZsh{}\PYZsh{}\PYZsh{}\PYZsh{}\PYZsh{}\PYZsh{}\PYZsh{}\PYZsh{}\PYZsh{}\PYZsh{}\PYZsh{}\PYZsh{}\PYZsh{}\PYZsh{}\PYZsh{}\PYZsh{}\PYZsh{}\PYZsh{}\PYZsh{}\PYZsh{}\PYZsh{}\PYZsh{}\PYZsh{}\PYZsh{}\PYZsh{}\PYZsh{}\PYZsh{}\PYZsh{}\PYZsh{}\PYZsh{}\PYZsh{}\PYZsh{}\PYZsh{}\PYZsh{}\PYZsh{}\PYZsh{}\PYZsh{}\PYZsh{}\PYZsh{}\PYZsh{}\PYZsh{}\PYZsh{}\PYZsh{}\PYZsh{}\PYZsh{}\PYZsh{}\PYZsh{}\PYZsh{}\PYZsh{}\PYZsh{}\PYZsh{}\PYZsh{}\PYZsh{}\PYZsh{}\PYZsh{}\PYZsh{}\PYZsh{}\PYZsh{}\PYZsh{}\PYZsh{}\PYZsh{}\PYZsh{}\PYZsh{}\PYZsh{}}
\PY{c+c1}{\PYZsh{}\PYZsh{}\PYZsh{}\PYZsh{}\PYZsh{}\PYZsh{}\PYZsh{}\PYZsh{}\PYZsh{}\PYZsh{}\PYZsh{}\PYZsh{}\PYZsh{}\PYZsh{}\PYZsh{}\PYZsh{}\PYZsh{}\PYZsh{}\PYZsh{}\PYZsh{}\PYZsh{}\PYZsh{}\PYZsh{}\PYZsh{}\PYZsh{}\PYZsh{}\PYZsh{}\PYZsh{}\PYZsh{}\PYZsh{}\PYZsh{}\PYZsh{}\PYZsh{}\PYZsh{}\PYZsh{}\PYZsh{}\PYZsh{}\PYZsh{}\PYZsh{}\PYZsh{}\PYZsh{}\PYZsh{}\PYZsh{}\PYZsh{}\PYZsh{}\PYZsh{}\PYZsh{}\PYZsh{}\PYZsh{}\PYZsh{}\PYZsh{}\PYZsh{}\PYZsh{}\PYZsh{}\PYZsh{}\PYZsh{}\PYZsh{}\PYZsh{}\PYZsh{}\PYZsh{}\PYZsh{}\PYZsh{}\PYZsh{}\PYZsh{}\PYZsh{}\PYZsh{}\PYZsh{}\PYZsh{}\PYZsh{}\PYZsh{}\PYZsh{}\PYZsh{}\PYZsh{}\PYZsh{}\PYZsh{}\PYZsh{}\PYZsh{}\PYZsh{}\PYZsh{}\PYZsh{}\PYZsh{}\PYZsh{}\PYZsh{}\PYZsh{}\PYZsh{}\PYZsh{}\PYZsh{}\PYZsh{}}
\PY{c+c1}{\PYZsh{}\PYZsh{}\PYZsh{}\PYZsh{}\PYZsh{}\PYZsh{}\PYZsh{}\PYZsh{}\PYZsh{}\PYZsh{}\PYZsh{}\PYZsh{}\PYZsh{}\PYZsh{}\PYZsh{}\PYZsh{}\PYZsh{}\PYZsh{}\PYZsh{}\PYZsh{}\PYZsh{}\PYZsh{}\PYZsh{}\PYZsh{}\PYZsh{}\PYZsh{}\PYZsh{}\PYZsh{}\PYZsh{}\PYZsh{}\PYZsh{}\PYZsh{}\PYZsh{}\PYZsh{} preamble and set\PYZhy{}up \PYZsh{}\PYZsh{}\PYZsh{}\PYZsh{}\PYZsh{}\PYZsh{}\PYZsh{}\PYZsh{}\PYZsh{}\PYZsh{}\PYZsh{}\PYZsh{}\PYZsh{}\PYZsh{}\PYZsh{}\PYZsh{}\PYZsh{}\PYZsh{}\PYZsh{}\PYZsh{}\PYZsh{}\PYZsh{}\PYZsh{}\PYZsh{}\PYZsh{}\PYZsh{}\PYZsh{}\PYZsh{}\PYZsh{}\PYZsh{}\PYZsh{}\PYZsh{}\PYZsh{}}
\PY{c+c1}{\PYZsh{}\PYZsh{}\PYZsh{}\PYZsh{}\PYZsh{}\PYZsh{}\PYZsh{}\PYZsh{}\PYZsh{}\PYZsh{}\PYZsh{}\PYZsh{}\PYZsh{}\PYZsh{}\PYZsh{}\PYZsh{}\PYZsh{}\PYZsh{}\PYZsh{}\PYZsh{}\PYZsh{}\PYZsh{}\PYZsh{}\PYZsh{}\PYZsh{}\PYZsh{}\PYZsh{}\PYZsh{}\PYZsh{}\PYZsh{}\PYZsh{}\PYZsh{}\PYZsh{}\PYZsh{}\PYZsh{}\PYZsh{}\PYZsh{}\PYZsh{}\PYZsh{}\PYZsh{}\PYZsh{}\PYZsh{}\PYZsh{}\PYZsh{}\PYZsh{}\PYZsh{}\PYZsh{}\PYZsh{}\PYZsh{}\PYZsh{}\PYZsh{}\PYZsh{}\PYZsh{}\PYZsh{}\PYZsh{}\PYZsh{}\PYZsh{}\PYZsh{}\PYZsh{}\PYZsh{}\PYZsh{}\PYZsh{}\PYZsh{}\PYZsh{}\PYZsh{}\PYZsh{}\PYZsh{}\PYZsh{}\PYZsh{}\PYZsh{}\PYZsh{}\PYZsh{}\PYZsh{}\PYZsh{}\PYZsh{}\PYZsh{}\PYZsh{}\PYZsh{}\PYZsh{}\PYZsh{}\PYZsh{}\PYZsh{}\PYZsh{}\PYZsh{}\PYZsh{}\PYZsh{}\PYZsh{}\PYZsh{}}
\PY{c+c1}{\PYZsh{}\PYZsh{}\PYZsh{}\PYZsh{}\PYZsh{}\PYZsh{}\PYZsh{}\PYZsh{}\PYZsh{}\PYZsh{}\PYZsh{}\PYZsh{}\PYZsh{}\PYZsh{}\PYZsh{}\PYZsh{}\PYZsh{}\PYZsh{}\PYZsh{}\PYZsh{}\PYZsh{}\PYZsh{}\PYZsh{}\PYZsh{}\PYZsh{}\PYZsh{}\PYZsh{}\PYZsh{}\PYZsh{}\PYZsh{}\PYZsh{}\PYZsh{}\PYZsh{}\PYZsh{}\PYZsh{}\PYZsh{}\PYZsh{}\PYZsh{}\PYZsh{}\PYZsh{}\PYZsh{}\PYZsh{}\PYZsh{}\PYZsh{}\PYZsh{}\PYZsh{}\PYZsh{}\PYZsh{}\PYZsh{}\PYZsh{}\PYZsh{}\PYZsh{}\PYZsh{}\PYZsh{}\PYZsh{}\PYZsh{}\PYZsh{}\PYZsh{}\PYZsh{}\PYZsh{}\PYZsh{}\PYZsh{}\PYZsh{}\PYZsh{}\PYZsh{}\PYZsh{}\PYZsh{}\PYZsh{}\PYZsh{}\PYZsh{}\PYZsh{}\PYZsh{}\PYZsh{}\PYZsh{}\PYZsh{}\PYZsh{}\PYZsh{}\PYZsh{}\PYZsh{}\PYZsh{}\PYZsh{}\PYZsh{}\PYZsh{}\PYZsh{}\PYZsh{}\PYZsh{}\PYZsh{}\PYZsh{}}

\PY{c+c1}{\PYZsh{} required to install github packages}
\PY{c+c1}{\PYZsh{} see http://cran.r\PYZhy{}project.org/web/packages/devtools/README.html}
\PY{k+kn}{library}\PY{p}{(}devtools\PY{p}{)}

\PY{c+c1}{\PYZsh{}install CLSA package}
devtools\PY{o}{::}install\PYZus{}github\PY{p}{(}\PY{l+s}{\PYZsq{}}\PY{l+s}{tystan/clsa\PYZsq{}}\PY{p}{)}
\PY{k+kn}{library}\PY{p}{(}clsa\PY{p}{)}
\PY{c+c1}{\PYZsh{} documentation}
\PY{o}{?}clsa\PYZus{}min

\PY{c+c1}{\PYZsh{} latex formatted table output}
\PY{k+kn}{library}\PY{p}{(}xtable\PY{p}{)}

\PY{c+c1}{\PYZsh{} function to create x values (m/z values)}
get\PYZus{}x\PYZus{}coords\PY{o}{\PYZlt{}\PYZhy{}}\PY{k+kr}{function}\PY{p}{(}n\PY{p}{)} \PY{k+kr}{return}\PY{p}{(}\PY{k+kp}{sort}\PY{p}{(}rbeta\PY{p}{(}n\PY{p}{,}\PY{l+m}{1}\PY{p}{,}\PY{l+m}{3}\PY{p}{)}\PY{p}{)}\PY{p}{)}
\PY{c+c1}{\PYZsh{} function to create intensity values}
get\PYZus{}f\PYZus{}signal\PY{o}{\PYZlt{}\PYZhy{}}\PY{k+kr}{function}\PY{p}{(}n\PY{p}{)} \PY{k+kr}{return}\PY{p}{(}rchisq\PY{p}{(}n\PY{p}{,}\PY{l+m}{10}\PY{p}{)}\PY{p}{)}

\PY{c+c1}{\PYZsh{}\PYZsh{}\PYZsh{}\PYZsh{}\PYZsh{}\PYZsh{}\PYZsh{}\PYZsh{}\PYZsh{}\PYZsh{}\PYZsh{}\PYZsh{}\PYZsh{}\PYZsh{}\PYZsh{}\PYZsh{}\PYZsh{}\PYZsh{}\PYZsh{}\PYZsh{}\PYZsh{}\PYZsh{}\PYZsh{}\PYZsh{}\PYZsh{}\PYZsh{}\PYZsh{}\PYZsh{}\PYZsh{}\PYZsh{}\PYZsh{}\PYZsh{}\PYZsh{}\PYZsh{}\PYZsh{}\PYZsh{}\PYZsh{}\PYZsh{}\PYZsh{}\PYZsh{}\PYZsh{}\PYZsh{}\PYZsh{}\PYZsh{}\PYZsh{}\PYZsh{}\PYZsh{}\PYZsh{}\PYZsh{}\PYZsh{}\PYZsh{}\PYZsh{}\PYZsh{}\PYZsh{}\PYZsh{}\PYZsh{}\PYZsh{}\PYZsh{}\PYZsh{}\PYZsh{}\PYZsh{}\PYZsh{}\PYZsh{}\PYZsh{}\PYZsh{}\PYZsh{}\PYZsh{}\PYZsh{}\PYZsh{}\PYZsh{}\PYZsh{}\PYZsh{}\PYZsh{}\PYZsh{}\PYZsh{}\PYZsh{}\PYZsh{}\PYZsh{}\PYZsh{}\PYZsh{}\PYZsh{}\PYZsh{}\PYZsh{}\PYZsh{}\PYZsh{}\PYZsh{}\PYZsh{}\PYZsh{}}
\PY{c+c1}{\PYZsh{}\PYZsh{}\PYZsh{}\PYZsh{}\PYZsh{}\PYZsh{}\PYZsh{}\PYZsh{}\PYZsh{}\PYZsh{}\PYZsh{}\PYZsh{}\PYZsh{}\PYZsh{}\PYZsh{}\PYZsh{}\PYZsh{}\PYZsh{}\PYZsh{}\PYZsh{}\PYZsh{}\PYZsh{}\PYZsh{}\PYZsh{}\PYZsh{}\PYZsh{}\PYZsh{}\PYZsh{}\PYZsh{}\PYZsh{}\PYZsh{}\PYZsh{}\PYZsh{}\PYZsh{}\PYZsh{}\PYZsh{}\PYZsh{}\PYZsh{}\PYZsh{}\PYZsh{}\PYZsh{}\PYZsh{}\PYZsh{}\PYZsh{}\PYZsh{}\PYZsh{}\PYZsh{}\PYZsh{}\PYZsh{}\PYZsh{}\PYZsh{}\PYZsh{}\PYZsh{}\PYZsh{}\PYZsh{}\PYZsh{}\PYZsh{}\PYZsh{}\PYZsh{}\PYZsh{}\PYZsh{}\PYZsh{}\PYZsh{}\PYZsh{}\PYZsh{}\PYZsh{}\PYZsh{}\PYZsh{}\PYZsh{}\PYZsh{}\PYZsh{}\PYZsh{}\PYZsh{}\PYZsh{}\PYZsh{}\PYZsh{}\PYZsh{}\PYZsh{}\PYZsh{}\PYZsh{}\PYZsh{}\PYZsh{}\PYZsh{}\PYZsh{}\PYZsh{}\PYZsh{}\PYZsh{}\PYZsh{}}
\PY{c+c1}{\PYZsh{}\PYZsh{}\PYZsh{}\PYZsh{}\PYZsh{}\PYZsh{}\PYZsh{}\PYZsh{}\PYZsh{}\PYZsh{}\PYZsh{}\PYZsh{}\PYZsh{}\PYZsh{}\PYZsh{}\PYZsh{}\PYZsh{}\PYZsh{}\PYZsh{}\PYZsh{}\PYZsh{}\PYZsh{}\PYZsh{}\PYZsh{}\PYZsh{}\PYZsh{}\PYZsh{}\PYZsh{}\PYZsh{}\PYZsh{}\PYZsh{}\PYZsh{}\PYZsh{}\PYZsh{} testing comp times \PYZsh{}\PYZsh{}\PYZsh{}\PYZsh{}\PYZsh{}\PYZsh{}\PYZsh{}\PYZsh{}\PYZsh{}\PYZsh{}\PYZsh{}\PYZsh{}\PYZsh{}\PYZsh{}\PYZsh{}\PYZsh{}\PYZsh{}\PYZsh{}\PYZsh{}\PYZsh{}\PYZsh{}\PYZsh{}\PYZsh{}\PYZsh{}\PYZsh{}\PYZsh{}\PYZsh{}\PYZsh{}\PYZsh{}\PYZsh{}\PYZsh{}\PYZsh{}\PYZsh{}\PYZsh{}}
\PY{c+c1}{\PYZsh{}\PYZsh{}\PYZsh{}\PYZsh{}\PYZsh{}\PYZsh{}\PYZsh{}\PYZsh{}\PYZsh{}\PYZsh{}\PYZsh{}\PYZsh{}\PYZsh{}\PYZsh{}\PYZsh{}\PYZsh{}\PYZsh{}\PYZsh{}\PYZsh{}\PYZsh{}\PYZsh{}\PYZsh{}\PYZsh{}\PYZsh{}\PYZsh{}\PYZsh{}\PYZsh{}\PYZsh{}\PYZsh{}\PYZsh{}\PYZsh{}\PYZsh{}\PYZsh{}\PYZsh{}\PYZsh{}\PYZsh{}\PYZsh{}\PYZsh{}\PYZsh{}\PYZsh{}\PYZsh{}\PYZsh{}\PYZsh{}\PYZsh{}\PYZsh{}\PYZsh{}\PYZsh{}\PYZsh{}\PYZsh{}\PYZsh{}\PYZsh{}\PYZsh{}\PYZsh{}\PYZsh{}\PYZsh{}\PYZsh{}\PYZsh{}\PYZsh{}\PYZsh{}\PYZsh{}\PYZsh{}\PYZsh{}\PYZsh{}\PYZsh{}\PYZsh{}\PYZsh{}\PYZsh{}\PYZsh{}\PYZsh{}\PYZsh{}\PYZsh{}\PYZsh{}\PYZsh{}\PYZsh{}\PYZsh{}\PYZsh{}\PYZsh{}\PYZsh{}\PYZsh{}\PYZsh{}\PYZsh{}\PYZsh{}\PYZsh{}\PYZsh{}\PYZsh{}\PYZsh{}\PYZsh{}\PYZsh{}}
\PY{c+c1}{\PYZsh{}\PYZsh{}\PYZsh{}\PYZsh{}\PYZsh{}\PYZsh{}\PYZsh{}\PYZsh{}\PYZsh{}\PYZsh{}\PYZsh{}\PYZsh{}\PYZsh{}\PYZsh{}\PYZsh{}\PYZsh{}\PYZsh{}\PYZsh{}\PYZsh{}\PYZsh{}\PYZsh{}\PYZsh{}\PYZsh{}\PYZsh{}\PYZsh{}\PYZsh{}\PYZsh{}\PYZsh{}\PYZsh{}\PYZsh{}\PYZsh{}\PYZsh{}\PYZsh{}\PYZsh{}\PYZsh{}\PYZsh{}\PYZsh{}\PYZsh{}\PYZsh{}\PYZsh{}\PYZsh{}\PYZsh{}\PYZsh{}\PYZsh{}\PYZsh{}\PYZsh{}\PYZsh{}\PYZsh{}\PYZsh{}\PYZsh{}\PYZsh{}\PYZsh{}\PYZsh{}\PYZsh{}\PYZsh{}\PYZsh{}\PYZsh{}\PYZsh{}\PYZsh{}\PYZsh{}\PYZsh{}\PYZsh{}\PYZsh{}\PYZsh{}\PYZsh{}\PYZsh{}\PYZsh{}\PYZsh{}\PYZsh{}\PYZsh{}\PYZsh{}\PYZsh{}\PYZsh{}\PYZsh{}\PYZsh{}\PYZsh{}\PYZsh{}\PYZsh{}\PYZsh{}\PYZsh{}\PYZsh{}\PYZsh{}\PYZsh{}\PYZsh{}\PYZsh{}\PYZsh{}\PYZsh{}\PYZsh{}}

\PY{c+c1}{\PYZsh{} creating a data.frame with the following columns:}
\PY{c+c1}{\PYZsh{} * n: the number of m/z points}
\PY{c+c1}{\PYZsh{} * win: the window sizes}
\PY{c+c1}{\PYZsh{} * time\PYZus{}naiv: using the naive alg \PYZhy{}\PYZhy{} time taken for the row\PYZsq{}s `n\PYZsq{} and `win\PYZsq{} values }
\PY{c+c1}{\PYZsh{} * time\PYZus{}clsa: using the CLSA \PYZhy{}\PYZhy{} time taken for the row\PYZsq{}s `n\PYZsq{} and `win\PYZsq{} values}

\PY{c+c1}{\PYZsh{} ranges of dataset size and window size:}
n\PYZus{}rng\PY{o}{\PYZlt{}\PYZhy{}}\PY{k+kp}{seq}\PY{p}{(}\PY{l+m}{1e4}\PY{p}{,}\PY{l+m}{1e5}\PY{p}{,}by\PY{o}{=}\PY{l+m}{1e4}\PY{p}{)}
win\PYZus{}rng\PY{o}{\PYZlt{}\PYZhy{}}\PY{k+kt}{c}\PY{p}{(}\PY{l+m}{0.005}\PY{p}{,}\PY{l+m}{0.01}\PY{p}{,}\PY{l+m}{0.02}\PY{p}{,}\PY{l+m}{0.05}\PY{p}{,}\PY{l+m}{0.1}\PY{p}{,}\PY{l+m}{0.2}\PY{p}{)}
\PY{c+c1}{\PYZsh{} these times will be updated}
time\PYZus{}naiv\PY{o}{\PYZlt{}\PYZhy{}}\PY{l+m}{0}
time\PYZus{}clsa\PY{o}{\PYZlt{}\PYZhy{}}\PY{l+m}{0}
\PY{c+c1}{\PYZsh{} enumerate all n and win combinations}
time\PYZus{}df\PY{o}{\PYZlt{}\PYZhy{}}\PY{k+kp}{as.data.frame}\PY{p}{(}
	\PY{k+kp}{expand.grid}\PY{p}{(}
		 n\PY{o}{=}n\PYZus{}rng
		\PY{p}{,}win\PY{o}{=}win\PYZus{}rng
		\PY{p}{,}time\PYZus{}naiv\PY{o}{=}time\PYZus{}naiv
		\PY{p}{,}time\PYZus{}clsa\PY{o}{=}time\PYZus{}clsa
	\PY{p}{)}
\PY{p}{)}
\PY{p}{(}n\PYZus{}df\PY{o}{\PYZlt{}\PYZhy{}}\PY{k+kp}{nrow}\PY{p}{(}time\PYZus{}df\PY{p}{)}\PY{p}{)} \PY{c+c1}{\PYZsh{} should be 6x10=60}
time\PYZus{}df

\PY{c+c1}{\PYZsh{} test each n and win combination 20 times, }
\PY{c+c1}{\PYZsh{} i.e., each dataset has 20 \PYZdq{}spectra\PYZdq{}}
n\PYZus{}reps\PY{o}{\PYZlt{}\PYZhy{}}\PY{l+m}{20} 
\PY{k+kp}{set.seed}\PY{p}{(}\PY{l+m}{12345}\PY{p}{)} \PY{c+c1}{\PYZsh{} make reproducible}

\PY{c+c1}{\PYZsh{} now iterate over rows of the data.frame for each \PYZdq{}spectrum\PYZdq{},}
\PY{c+c1}{\PYZsh{} and time the computation}
\PY{k+kr}{for}\PY{p}{(}j \PY{k+kr}{in} \PY{l+m}{1}\PY{o}{:}n\PYZus{}reps\PY{p}{)}
\PY{p}{\PYZob{}}
	\PY{k+kr}{for}\PY{p}{(}i \PY{k+kr}{in} \PY{l+m}{1}\PY{o}{:}n\PYZus{}df\PY{p}{)}
	\PY{p}{\PYZob{}}

		\PY{c+c1}{\PYZsh{} get n and win combination}
		n\PY{o}{\PYZlt{}\PYZhy{}}time\PYZus{}df\PY{o}{\PYZdl{}}n\PY{p}{[}i\PY{p}{]}
		this\PYZus{}win\PY{o}{\PYZlt{}\PYZhy{}}time\PYZus{}df\PY{o}{\PYZdl{}}win\PY{p}{[}i\PY{p}{]}
		\PY{c+c1}{\PYZsh{} update console of progress}
		\PY{k+kp}{cat}\PY{p}{(}\PY{l+s}{\PYZdq{}}\PY{l+s}{::: Iteration\PYZdq{}}\PY{p}{,}\PY{p}{(}j\PY{l+m}{\PYZhy{}1}\PY{p}{)}\PY{o}{*}n\PYZus{}df\PY{o}{+}i\PY{p}{,}\PY{l+s}{\PYZdq{}}\PY{l+s}{of\PYZdq{}}\PY{p}{,}n\PYZus{}df\PY{o}{*}n\PYZus{}reps\PY{p}{,}\PY{l+s}{\PYZdq{}}\PY{l+s}{::: \PYZdq{}}\PY{p}{)}
		\PY{k+kp}{cat}\PY{p}{(}\PY{l+s}{\PYZdq{}}\PY{l+s}{n =\PYZdq{}}\PY{p}{,}n\PY{p}{,}\PY{l+s}{\PYZdq{}}\PY{l+s}{and window size = \PYZdq{}}\PY{p}{,}this\PYZus{}win\PY{p}{,}\PY{l+s}{\PYZdq{}}\PY{l+s}{:::\PYZbs{}n\PYZdq{}}\PY{p}{)}
		\PY{c+c1}{\PYZsh{} randomly generate x and f}
		x\PY{o}{\PYZlt{}\PYZhy{}}get\PYZus{}x\PYZus{}coords\PY{p}{(}n\PY{p}{)}
		f\PY{o}{\PYZlt{}\PYZhy{}}get\PYZus{}f\PYZus{}signal\PY{p}{(}n\PY{p}{)}

		\PY{c+c1}{\PYZsh{} time the computations, add to previous \PYZdq{}spectra\PYZdq{} times}
		time\PYZus{}df\PY{o}{\PYZdl{}}time\PYZus{}clsa\PY{p}{[}i\PY{p}{]}\PY{o}{\PYZlt{}\PYZhy{}}time\PYZus{}df\PY{o}{\PYZdl{}}time\PYZus{}clsa\PY{p}{[}i\PY{p}{]}\PY{o}{+}
			\PY{k+kp}{system.time}\PY{p}{(}a\PY{o}{\PYZlt{}\PYZhy{}}clsa\PYZus{}max\PY{p}{(}x\PY{p}{,}clsa\PYZus{}min\PY{p}{(}x\PY{p}{,}f\PY{p}{,}this\PYZus{}win\PY{p}{)}\PY{p}{,}this\PYZus{}win\PY{p}{)}\PY{p}{)}\PY{p}{[}\PY{l+m}{3}\PY{p}{]}
		time\PYZus{}df\PY{o}{\PYZdl{}}time\PYZus{}naiv\PY{p}{[}i\PY{p}{]}\PY{o}{\PYZlt{}\PYZhy{}}time\PYZus{}df\PY{o}{\PYZdl{}}time\PYZus{}naiv\PY{p}{[}i\PY{p}{]}\PY{o}{+}
			\PY{k+kp}{system.time}\PY{p}{(}b\PY{o}{\PYZlt{}\PYZhy{}}naiv\PYZus{}max\PY{p}{(}x\PY{p}{,}naiv\PYZus{}min\PY{p}{(}x\PY{p}{,}f\PY{p}{,}this\PYZus{}win\PY{p}{)}\PY{p}{,}this\PYZus{}win\PY{p}{)}\PY{p}{)}\PY{p}{[}\PY{l+m}{3}\PY{p}{]}

		\PY{c+c1}{\PYZsh{} if the results are not equal between the naiv and CLSA we have }
		\PY{c+c1}{\PYZsh{} a problem; ABORT!}
		\PY{k+kr}{if}\PY{p}{(}\PY{o}{!}\PY{k+kp}{all}\PY{p}{(}a\PY{o}{==}b\PY{p}{)}\PY{p}{)} \PY{k+kr}{break}\PY{p}{;} 
	\PY{p}{\PYZcb{}}
\PY{p}{\PYZcb{}}
time\PYZus{}df

\PY{c+c1}{\PYZsh{}\PYZsh{}\PYZsh{}\PYZsh{}\PYZsh{}\PYZsh{}\PYZsh{}\PYZsh{}\PYZsh{}\PYZsh{}\PYZsh{}\PYZsh{}\PYZsh{}\PYZsh{}\PYZsh{}\PYZsh{}\PYZsh{}\PYZsh{}\PYZsh{}\PYZsh{}\PYZsh{}\PYZsh{}\PYZsh{}\PYZsh{}\PYZsh{}\PYZsh{}\PYZsh{}\PYZsh{}\PYZsh{}\PYZsh{}\PYZsh{}\PYZsh{}\PYZsh{}\PYZsh{}\PYZsh{}\PYZsh{}\PYZsh{}\PYZsh{}\PYZsh{}\PYZsh{}\PYZsh{}\PYZsh{}\PYZsh{}\PYZsh{}\PYZsh{}\PYZsh{}\PYZsh{}\PYZsh{}\PYZsh{}\PYZsh{}\PYZsh{}\PYZsh{}\PYZsh{}\PYZsh{}\PYZsh{}\PYZsh{}\PYZsh{}\PYZsh{}\PYZsh{}\PYZsh{}\PYZsh{}\PYZsh{}\PYZsh{}\PYZsh{}\PYZsh{}\PYZsh{}\PYZsh{}\PYZsh{}\PYZsh{}\PYZsh{}\PYZsh{}\PYZsh{}\PYZsh{}\PYZsh{}\PYZsh{}\PYZsh{}\PYZsh{}\PYZsh{}\PYZsh{}\PYZsh{}\PYZsh{}\PYZsh{}\PYZsh{}\PYZsh{}\PYZsh{}\PYZsh{}\PYZsh{}\PYZsh{}}
\PY{c+c1}{\PYZsh{}\PYZsh{}\PYZsh{}\PYZsh{}\PYZsh{}\PYZsh{}\PYZsh{}\PYZsh{}\PYZsh{}\PYZsh{}\PYZsh{}\PYZsh{}\PYZsh{}\PYZsh{}\PYZsh{}\PYZsh{}\PYZsh{}\PYZsh{}\PYZsh{}\PYZsh{}\PYZsh{}\PYZsh{}\PYZsh{}\PYZsh{}\PYZsh{}\PYZsh{}\PYZsh{}\PYZsh{}\PYZsh{}\PYZsh{}\PYZsh{}\PYZsh{}\PYZsh{}\PYZsh{}\PYZsh{} table of results \PYZsh{}\PYZsh{}\PYZsh{}\PYZsh{}\PYZsh{}\PYZsh{}\PYZsh{}\PYZsh{}\PYZsh{}\PYZsh{}\PYZsh{}\PYZsh{}\PYZsh{}\PYZsh{}\PYZsh{}\PYZsh{}\PYZsh{}\PYZsh{}\PYZsh{}\PYZsh{}\PYZsh{}\PYZsh{}\PYZsh{}\PYZsh{}\PYZsh{}\PYZsh{}\PYZsh{}\PYZsh{}\PYZsh{}\PYZsh{}\PYZsh{}\PYZsh{}\PYZsh{}\PYZsh{}\PYZsh{}}
\PY{c+c1}{\PYZsh{}\PYZsh{}\PYZsh{}\PYZsh{}\PYZsh{}\PYZsh{}\PYZsh{}\PYZsh{}\PYZsh{}\PYZsh{}\PYZsh{}\PYZsh{}\PYZsh{}\PYZsh{}\PYZsh{}\PYZsh{}\PYZsh{}\PYZsh{}\PYZsh{}\PYZsh{}\PYZsh{}\PYZsh{}\PYZsh{}\PYZsh{}\PYZsh{}\PYZsh{}\PYZsh{}\PYZsh{}\PYZsh{}\PYZsh{}\PYZsh{}\PYZsh{}\PYZsh{}\PYZsh{}\PYZsh{}\PYZsh{}\PYZsh{}\PYZsh{}\PYZsh{}\PYZsh{}\PYZsh{}\PYZsh{}\PYZsh{}\PYZsh{}\PYZsh{}\PYZsh{}\PYZsh{}\PYZsh{}\PYZsh{}\PYZsh{}\PYZsh{}\PYZsh{}\PYZsh{}\PYZsh{}\PYZsh{}\PYZsh{}\PYZsh{}\PYZsh{}\PYZsh{}\PYZsh{}\PYZsh{}\PYZsh{}\PYZsh{}\PYZsh{}\PYZsh{}\PYZsh{}\PYZsh{}\PYZsh{}\PYZsh{}\PYZsh{}\PYZsh{}\PYZsh{}\PYZsh{}\PYZsh{}\PYZsh{}\PYZsh{}\PYZsh{}\PYZsh{}\PYZsh{}\PYZsh{}\PYZsh{}\PYZsh{}\PYZsh{}\PYZsh{}\PYZsh{}\PYZsh{}\PYZsh{}\PYZsh{}}

\PY{c+c1}{\PYZsh{} extract times for naiv and CLSA for printing}
time\PYZus{}naiv\PY{o}{\PYZlt{}\PYZhy{}}\PY{k+kt}{data.frame}\PY{p}{(}n\PY{o}{=}time\PYZus{}df\PY{o}{\PYZdl{}}n\PY{p}{,}win\PY{o}{=}time\PYZus{}df\PY{o}{\PYZdl{}}win\PY{p}{,}time\PY{o}{=}time\PYZus{}df\PY{o}{\PYZdl{}}time\PYZus{}naiv\PY{p}{,}func\PY{o}{=}\PY{l+s}{\PYZdq{}}\PY{l+s}{Naive\PYZdq{}}\PY{p}{)}
time\PYZus{}clsa\PY{o}{\PYZlt{}\PYZhy{}}\PY{k+kt}{data.frame}\PY{p}{(}n\PY{o}{=}time\PYZus{}df\PY{o}{\PYZdl{}}n\PY{p}{,}win\PY{o}{=}time\PYZus{}df\PY{o}{\PYZdl{}}win\PY{p}{,}time\PY{o}{=}time\PYZus{}df\PY{o}{\PYZdl{}}time\PYZus{}clsa\PY{p}{,}func\PY{o}{=}\PY{l+s}{\PYZdq{}}\PY{l+s}{CLSA\PYZdq{}}\PY{p}{)}

\PY{c+c1}{\PYZsh{} print the results!}
xtable\PY{p}{(}
	\PY{k+kp}{cbind}\PY{p}{(}
		xtabs\PY{p}{(}time \PY{o}{\PYZti{}} \PY{k+kp}{I}\PY{p}{(}n\PY{o}{/}\PY{l+m}{1e4}\PY{p}{)} \PY{o}{+} win\PY{p}{,} data \PY{o}{=} time\PYZus{}naiv\PY{p}{)}
		\PY{p}{,}\PY{k+kc}{NA}
		\PY{p}{,}xtabs\PY{p}{(}time \PY{o}{\PYZti{}} \PY{k+kp}{I}\PY{p}{(}n\PY{o}{/}\PY{l+m}{1e4}\PY{p}{)} \PY{o}{+} win\PY{p}{,} data \PY{o}{=} time\PYZus{}clsa\PY{p}{)}
	\PY{p}{)}
	\PY{p}{,}digits\PY{o}{=}\PY{l+m}{1}
\PY{p}{)}
\end{Verbatim}